\documentclass[lettersize,journal]{IEEEtran}
\usepackage[utf8]{inputenc}
\usepackage[english]{babel} 
\usepackage[linesnumbered]{algorithm2e}
\usepackage{amsmath,amsfonts, amssymb, amsthm} 
\usepackage{mathtools}
\usepackage{acronym}
\usepackage{longtable}
\usepackage{textcomp}
\usepackage{multirow}
\usepackage{makecell}
\usepackage{subfigure}
\usepackage{graphicx}
\usepackage{color}
\usepackage{epsfig}
\usepackage{url}
\usepackage{balance}
\usepackage{float}
\usepackage{comment}
\usepackage{pifont}
\usepackage{hyperref}
\usepackage{xr}
\usepackage{array}
\usepackage{todonotes}
\usepackage{bm}
\usepackage{tcolorbox}
\usepackage{aircraftshapes}
\usepackage{pgf-umlsd}
\usetikzlibrary{decorations.pathreplacing}

\usepackage[
	n,
	operators,
	advantage,
	sets,
	adversary,
	landau,
	probability,
	notions,	
	logic,
	ff,
	mm,
	primitives,
	events,
	complexity,
	asymptotics,
	keys]{cryptocode}

\newcolumntype{P}[1]{>{\centering\arraybackslash}p{#1}}
\newcommand{\proto}{$A^2RID$}

\theoremstyle{definition}
\newtheorem{definition}{Definition}[section]

\acrodef{FAA}{Federal Avionics Administration}
\acrodef{ADS-B}{Automatic Dependent Surveillance - Broadcast}
\acrodef{ECC}{Elliptic Curve Cryptography}
\acrodef{GSM}{Global System for Mobile Communications}
\acrodef{GPS}{Global Positioning System}
\acrodef{GNSS}{Global Navigation Satellite System}
\acrodef{MTU}{Maximum Transmission Unit}
\acrodef{OS}{Operative System}
\acrodef{OSN}{Online Social Network}
\acrodef{P2P}{Peer-to-Peer}
\acrodef{PDU}{Protocol Data Unit}
\acrodef{PRNG}{Pseudo Random Number Generator}
\acrodef{TLS}{Transport Layer Security}
\acrodef{TTP}{Trusted Third-Party}
\acrodef{DSig}{Digital Signature}
\acrodef{PKE}{Public Key Encryption}
\acrodef{SPS-EQ}{Structure-Preserving Signatures on Equivalence Classes}
\acrodef{NIZKP}{Non-Interactive Zero Knowledge Proofs}
\acrodef{SoK}{Signature of Knowledge}
\acrodef{UA}{Unmanned Aircraft}
\acrodef{DRIP}{Drone Remote Identification Protocol}
\acrodef{RID}{Remote Identification}
\acrodef{USS}{Unmanned Service Supplier}
\acrodef{PbIR}{Public Information Registry}
\acrodef{CI}{Critical Infrastructure}
\acrodef{DoS}{Denial of Service}
\acrodef{GCS}{Ground Control Station}
\acrodef{VANET}{Vehicular Ad-Hoc Networks}
\acrodef{WG}{Working Group}
\acrodef{IETF}{Internet Engineering Task Force}
\acrodef{DDH}{Decisional Diffie–Hellman}

\begin{document}
\bstctlcite{IEEEexample:BSTcontrol}

\title{$A^{2}RID$ - Anonymous Direct Authentication and Remote Identification of Commercial Drones}

\author{Eva Wisse, Pietro Tedeschi, Savio Sciancalepore, and Roberto Di Pietro
\IEEEcompsocitemizethanks{
This is a personal copy of the authors. Not for redistribution. The final published version of the paper accepted in IEEE Internet of Things Journal is available through the IEEExplore Digital Library, with the DOI: 10.1109/JIOT.2023.3240477.\\\\
\IEEEcompsocthanksitem Eva Wisse is with the Eindhoven University of Technology (TU/e), Department of Mathematics and Computer Science, Eindhoven, Netherlands.
e-mail: e.m.c.wisse@student.tue.nl.
\IEEEcompsocthanksitem Pietro Tedeschi is with the Technology Innovation Institute, Autonomous Robotics Research Center, Abu Dhabi, United Arab Emirates.
e-mail: pietro.tedeschi@tii.ae
\IEEEcompsocthanksitem Savio Sciancalepore is with the Eindhoven University of Technology (TU/e), Department of Mathematics and Computer Science, Eindhoven, Netherlands.
e-mail: s.sciancalepore@tue.nl.
\IEEEcompsocthanksitem Roberto Di Pietro is with the Division of Information and Computing Technology (ICT), College of Science and Engineering (CSE), Hamad Bin Khalifa University (HBKU), Doha, Qatar.
e-mail: rdipietro@hbku.edu.qa.
}
}
\maketitle

\begin{abstract} 
    The recent worldwide introduction of RemoteID (RID) regulations forces all Unmanned Aircrafts (UAs), a.k.a. drones, to broadcast in plaintext on the wireless channel their identity and real-time location, for accounting and monitoring purposes. Although improving drones’ monitoring and situational awareness, the RID rule also generates significant privacy concerns for UAs’ operators, threatened by the ease of tracking of UAs and related confidentiality and privacy concerns connected with the broadcasting of plaintext identity information.\\
    In this paper, we propose $A^2RID$, a protocol suite for anonymous direct authentication and remote identification of heterogeneous commercial UAs. $A^2RID$ integrates and adapts protocols for anonymous message signing to work in the UA domain, coping with the constraints of commercial drones and the tight real-time requirements imposed by the RID regulation. Overall, the protocols in the $A^2RID$ suite allow a UA manufacturer to pick the configuration that best suits the capabilities and constraints of the drone, i.e., either a processing-intensive but memory-lightweight solution (namely, $CS-A^2RID$) or a computationally-friendly but memory-hungry approach (namely, $DS-A^2RID$). 
    Besides formally defining the protocols and formally proving their security in our setting, we also implement and test them on real heterogeneous hardware platforms, i.e., the Holybro X-500 and the ESPcopter, releasing open-source the produced code. For all the protocols, we demonstrated experimentally the capability of generating anonymous RemoteID messages well below the time bound of $1$ second required by RID, while at the same time having quite a limited impact on the energy budget of the drone.
\end{abstract}


\IEEEpeerreviewmaketitle

\section{Introduction}
\label{sec:intro}

\acp{UA}, a.k.a. drones, are gaining increasing momentum in Industry and Academia, thanks to the flexibility and enhanced mobility features they provide in several key application domains, e.g., surveillance, goods delivery, and search-and-rescue, to name a few~\cite{abualigah2021_sensors}. In addition, leading business forecasting companies estimate the drones' market to grow from USD $8.15$ billion in 2022 to USD $47.38$ billion by 2029, with a CAGR of approx. $28.58$\%, estimating more than $9.64$ million drones flying around by the same time~\cite{market},~\cite{market_2}.

Such large numbers motivated recent significant efforts by several regional authorities to integrate \acp{UA} within the local airspace, for traffic management and safety issues. In this context, the US-based \ac{FAA} was the first to take action, by introducing the \ac{RID} regulation~\cite{faa_page}. In a nutshell, \ac{RID} requires all \acp{UA} to broadcast on the wireless channel, from take-off to landing, information such as the identity and location of the \ac{UA}, with a minimum rate of $1$~msg/s (see Sec.~\ref{sec:remoteid} for more details). At the same time, initiatives similar to \ac{RID} are also planned in EU and China~\cite{remoteid_europe},~\cite{belwafi2022_access}.

Although solving traffic management and safety concerns, \ac{RID} regulations also introduce significant privacy issues~\cite{dronedj_privacy},~\cite{ainonline}. Indeed, by simply eavesdropping on the wireless channel, passive adversaries might easily get the unique identity of the \ac{UA}, its real-time location, and other sensitive information, such as the location of the related \ac{GCS}. Through a longer, fully passive and stealthy observation, the adversary can also track the drone during regular operations and infer more private information about their operators, such as the place where they live, the usual flight source, path, and destination, as well as the location of storage sites of large commercial delivery companies (e.g., in case of drone-based deliveries by goods distribution companies)~\cite{dji_app}. Such threats are also magnified by recent reports, documenting the leakage of drones identifiers on the public Internet~\cite{dji_dataLeak}. 
In this context, if the broadcast \ac{RID} messages were anonymous, \acp{UA} could protect the privacy of their operator(s), and make indiscriminate tracking much more challenging. At the same time, any solutions for anonymous remote identification should also guarantee the disclosure of the identity of possibly misbehaving \acp{UA}, i.e., when drones invade (accidentally or not) no-fly-zones. Moreover, such solutions should also cope with the limitations of \acp{UA}, mainly in terms of available processing and energy capabilities.

\textcolor{black}{At the time of this writing, very few scientific contributions investigated anonymous remote identification of \acp{UA} within the framework of the \ac{RID} regulation. In this context, a conference paper~\cite{tedeschi2021_acsac} of ours proposed \emph{ARID}, the first solution for anonymous remote identification of \acp{UA}. ARID allows \acp{UA} to broadcast anonymous \ac{RID} messages, where the long-term identity of the emitter is never revealed on the wireless channel.} At the same time, whenever the invasion of a no-fly area is detected, \ac{CI} operators might forward the received messages to a \ac{TTP}, such as the \ac{FAA}, to disclose the long-term identity of the misbehaving \ac{UA} and take action. However, ARID provides a brokered message authentication. Indeed, entities receiving ARID messages cannot directly and autonomously verify message authenticity, but they have to interact with the \ac{TTP} to verify that received messages are neither forged nor replayed. As a result, the deployment of ARID and its integration into the \ac{RID} framework might pose excessive management overhead on regulatory authorities. Moreover, being based on different entities, the networking architecture required by ARID does not match with current standardization activities, such as the ones carried out by the IETF WG \emph{drip}, working on the standardization of the components and the network architecture for the integration of \ac{RID} into national airspaces. 

{\bf Contribution.} In this paper, we make the following contributions.
\begin{itemize}
    \item We propose \proto\ (acronym for Anonymous direct Authentication and Remote IDentification), the first protocol suite for anonymous direct authentication and remote identification of heterogeneous commercial drones.
    \item Within the \proto\ protocol suite, we propose and define three protocols, namely: (i) $CS-A^2RID$, for high-end \acp{UA} equipped with regular processing and energy capabilities, capable of running pairing-based cryptography schemes on board; (ii) $DS-CCA2-A^2RID$, for \acp{UA} with low processing and energy availability, but equipped with large storage space; and, (iii) $DS-CPA-A^2RID$, for \acp{UA} characterized by severely constrained processing, storage, and energy availability.
    \item Through the protocols listed above, we provide a solution for the \acp{UA} to broadcast anonymous \ac{RID} messages, protecting their long-term unique identity from malicious eavesdroppers while being compliant with current \ac{RID} regulations. 
    \item For all the protocols above listed, we provide a rigorous protocol description within the network architecture for \ac{UA} remote identification defined by the IETF \ac{WG} \emph{drip}, as well as a formal security proof, via the well-known automated verification tool \pcalgostyle{ProVerif}.
    \item To show the viability of the proposed solution, we implemented the protocols in the \proto\ protocol suite on heterogeneous commercial \acp{UA}, i.e., the Holybro X500 and the ESPcopter, characterized by very different processing, storage, and energy capabilities. We also released the corresponding source code as open-source, to foster the reproducibility and re-usability of our code and results~\cite{arid2_code},~\cite{arid2_code_Eva}, as well as to stimulate further research in the field. 
    \item Finally, we also report the results of an extensive experimental performance assessment of our solutions when run on real heterogeneous hardware, demonstrating that it is possible to achieve anonymous remote identification of \acp{UA} in $\approx0.017$~seconds on the Holybro X500 and within $0.22$~seconds on the ESPcopter, i.e., well below the time limit of $1$~second imposed by the \ac{RID} regulation, even with severely constrained \acp{UA}. 
\end{itemize}

{\bf Roadmap.}
The rest of this paper is organized as follows. Sec.~\ref{sec:background} introduces preliminary notions, Sec.~\ref{sec:related} reviews the related work,  Sec.~\ref{sec:scenario_adv_model} outlines the scenario and the adversarial model considered in our work, Sec.~\ref{sec:proto} provides the details of our solution, Sec.~\ref{sec:security} discusses the security features offered by our solution, Sec.~\ref{sec:performance} provides a thorough performance evaluation, both via simulations and a real experimentation and, finally, Sec.~\ref{sec:conclusion} concludes the paper.

\section{Background and Preliminaries}
\label{sec:background}

In this section, we introduce background material that will be helpful for the sequel of the manuscript. Sec.~\ref{sec:remoteid} provides an overview of the \ac{RID} regulation, while Sec.~\ref{sec:crypto} summarizes cryptography techniques and notions used in this manuscript.

\subsection{RemoteID Specification}
\label{sec:remoteid}

The \ac{RID} rule was published first in April 2021 by the US-based \ac{FAA}, and it is set to become mandatory for all \acp{UA} from September 2022~\cite{faa_page}. According to the \ac{RID} specification, all \acp{UA}, almost independently from their weight and usage, should broadcast on the wireless channel the following information: (i) unique identifier, (ii) timestamp, (iii) current location, (iv) current speed, (v) location of any \ac{GCS}, and finally, (vi) emergency status. Such information should be broadcasted in plaintext, from take-off to landing time, with a minimum rate of one message per second. The rule also suggests the adoption of the WiFi standard for messages broadcasting, due to its reasonable range and widespread adoption. Besides the \emph{broadcast mode}, \ac{RID} also defines a \emph{unicast} mode, where \acp{UA} might be available on a given port to answer requests about their identity and location. At the same time, \ac{RID} does not force \acp{UA} to integrate an Internet connection, but only to feature a module for the broadcast of wireless messages. When such a module is unavailable, the manufacturer can provide dedicated external modules after the deployment to make \acp{UA} compliant with \ac{RID}.

Overall, the aim of the \ac{RID} rule is to set a framework for accountability of UA operations, as well as identification of the owner of any flying \acp{UA}. However, note that network security issues connected with the integration of \ac{RID} are specifically not addressed in the FAA rule.
Finally, it is worth noting that the overall problem of \acp{UA} remote identification goes beyond the US borders, and also other geographical airspaces such as the EU, Russia, and China are taking initiatives toward regulating drones' flights~\cite{remoteid_europe},~\cite{belwafi2022_access}.

\subsection{Cryptography Techniques and Notions}
\label{sec:crypto}

In this section, we recall as preliminaries the main building blocks used throughout the manuscript. \\\\
\noindent
\textbf{Public Key Encryption. } \label{sec:pke}
\ac{PKE} schemes allow to encrypt a message $M$ using the public key of the recipient $pk$ to a ciphertext $c$, such that the recipient only, in possession of the corresponding secret key $sk$, can decrypt the ciphertext and recover the plaintext message $M$.
\begin{definition}
A public key encryption algorithm $PKE$ consists of the following algorithms:\\
$PKE.KGen(1^k)$: on input a security parameter $k$, it outputs a secret decryption key $sk$ and a public encryption key $pk$.\\
$PKE.Enc(pk,m)$: on input a plaintext message $m$ and a public key $pk$, it outputs a ciphertext $c$. \\
$PKE.Dec(sk,c)$: on input a ciphertext $c$ and a secret decryption key $sk$, it outputs the corresponding plaintext $m$.
\end{definition}

Although any public-key encryption scheme can be used, in this manuscript, we use the well-known Elliptic Curve Integrated Encryption Scheme (ECIES) scheme~\cite{certicom2009}.\\\\

\noindent
\textbf{Decisional Diffie-Hellman (DDH).}
The \ac{DDH} is an assumption commonly used in cryptography on the computational hardness of solving discrete logarithms problems in cyclic groups. Such an assumption is at the roots of the security of many protocols, including Cramer–Shoup cryptosystems (see below). Assume $\mathbb{G}$ is a cyclic group of order $q$, with generator $g$, and $a, b, c$ are random values $\in \mathbb{Z}_q$. According to the DDH assumption, given the distributions $\langle g^a, g^b, g^{ab}\rangle$ and $\langle g^a, g^b, g^c\rangle$, they are computationally indistinguishable in the security parameter $n=log(q)$~\cite{boneh1998ddh}.
\\\\
\noindent
\textbf{Cramer-Shoup Cryptosystem}. Assume $\mathbb{G}$ is a cyclic group of prime order $q$, where $q$ is large, $m$ is a plaintext message encoded as an element of $\mathbb{G}$, and $H$ a universal family of one-way hash functions mapping bit-strings into elements of $\mathbb{Z}_q$~\cite{cramershoup_iacr}. 
\begin{definition}
A Cramer-Shoup public key encryption algorithm $CSC$ consists of the following algorithms:\\
$CSC.KGen(\mathbb{G}, \mathbb{Z}_q)$: on input a group $\mathbb{G}$ of prime order $q$, it generates random elements $g_1,g_2 \in \mathbb{G}$, and $x_1,x_2,y_1,y_2,z\in \mathbb{Z}_q$. Next, it computes the elements $c=g_1^{x_1}g_2^{x_2}, d=g_1^{y_1}g_2^{y_2},h=g_1^z$. The generated public key is the tuple $pk=(g_1, g_2, c, d, h, H)$, and the private
key is $sk=(x_1, x_2, y_1, y_2, z)$.\\
$CSC.Enc(pk,m,r)$: on input a plaintext message $m\in\mathbb{G}$, a public key $pk$, and $r\in\mathbb{Z}_q$ it outputs the ciphertext $c=(u_1,u_2,e,\psi)$, where $u_1=g_1^r, u_2=g_2^r, e=h^rm, \alpha=H(u_1,u_2,e), \psi=c^rd^{r\alpha}$.\\
$CSC.Dec(sk,c)$: on input a ciphertext $c=(u_1,u_2,e,\psi)$ and a secret key $sk=(x_1, x_2, y_1, y_2, z)$, the decryption algorithm computes $\alpha=H(u_1,u_2,e)$ and tests if $u_1^{x_1+y_1\alpha}u_2^{x_2+y_2\alpha} \stackrel{?}{=} \psi$. If this condition is verified, the algorithm outputs the plaintext $m=\frac{e}{u_1^z}$; otherwise it outputs $\bot$.
\end{definition}

\noindent
\textbf{Digital Signature Schemes. } \label{sec:signature}
\ac{DSig} schemes allow a sender to produce a signed value $\sigma$ for a message $m$, demonstrating to be the actual sender of the message.
\begin{definition}
A digital signature scheme $DSig$ consists of the following algorithms:\\
$DSig.KGen(1^K)$: on input a security parameter $k$, it outputs a secret signing key $sk$ and a public verification key $pk$.\\
$DSig.Sign(sk,m)$: on input a message $m$ and a signing key $sk$, it outputs a signature $\sigma$.\\
$DSig.Vrf(pk,m,\sigma)$: on input a message $m$, a public key $pk$, and a signature $\sigma$, it outputs a bit $b \in {0,1}$, where $0$ indicates that $\sigma$ is not verified and $1$ indicates that $\sigma$ is verified.
\end{definition}

Although any DSig scheme on elliptic curves can be used, in this manuscript, we use the scheme proposed by Boneh, Lynn, and Shacham~\cite{boneh2003_tact}.

\noindent
\textbf{Bilinear Pairings. }
Let $\mathbb{G}$ and $\mathbb{G}_T$ be multiplicative groups of prime order $q$, and $g$ be a generator of $\mathbb{G}$. A map $\hat{e}:\mathbb{G}\times\mathbb{G} \rightarrow \mathbb{G}_T$ is called a bilinear map if it satisfies the following properties:
\begin{enumerate}
    \item Bilinearity: $\hat{e}(g^\alpha, g^\beta)=\hat{e}(g, g)^{\alpha\beta}$ for all $\alpha, \beta \in \mathbb{Z_q^*}$.
    \item Non-degeneracy: There exist $\alpha,\beta\in\mathbb{G}$ such that $\hat{e}(\alpha, \beta)=1$.
    \item Computability: There exists an efficient algorithm to compute $\hat{e}(\alpha, \beta)$ for any $\alpha,\beta\in\mathbb{G}$.
\end{enumerate}

Several bilinear pairing types exist, i.e., Type-1 (symmetric), Type-2, Type-3, and Type-4. In this paper, we use Type-1 pairing ($\mathbb{G}_1 = \mathbb{G}_2$) and Type-3 pairing ($\mathbb{G}_1 \neq \mathbb{G}_2$ and absence of any computable isomorphism). Interested readers can refer to the article by the authors in~\cite{galbraith2008_dam} and~\cite{Chatterjee2010} for more details.

\noindent
\textbf{Structure-Preserving Signatures on Equivalence Classes. } 
\label{sec:sps_eq}
Structure-preserving signatures on equivalence classes allow, among other properties, to generate unlinkable message-signature pairs on elements of the same equivalence classes.
\begin{definition}
A structure-preserving signatures on equivalence classes scheme $S$ consists of the following algorithms.\\
$S.BGGen(1^K)$: on input a security parameter $k$, it outputs a bilinear group $BG$.\\
$S.KGen(BG,l)$: on input a bilinear group $BG$ and an integer $l$, it outputs a secret key $sk$ and a public key $pk$.\\
$S.Sign(m,sk)$: on input a message $m$ and a secret key $sk$, it outputs a signature $\sigma$.\\
$S.ChgRep(m,\sigma,\rho,pk)$: on input a message $m$, a signature $\sigma$ on $m$, a scalar $\rho$, and a public key $pk$, it outputs a message-signature pair $(M',\rho')$, being $M'=\rho \cdot M$.\\
$S.Vrf(m,\sigma,pk)$: on input a message $m$, a signature $\sigma$, and a public key $pk$, it outputs a bit $b \in {0,1}$, where $0$ indicates that $\sigma$ is not verified and $1$ indicates that $\sigma$ is verified.\\
$S.VKey(sk,pk)$: on input a secret key $sk$ and a public key $pk$, it outputs a bit $b \in {0,1}$, where $0$ indicates that the keys are not related to each other, while a $1$ indicates that they are related to each other.
\end{definition}

In this paper, we use the structure-preserving signatures on equivalence classes scheme reported in~\cite{derler2018_asiaccs}, and originally defined by the authors in~\cite{fuchsbauer2019_jcr}.\\\\
\noindent
\textbf{Non-Interactive Zero-Knowledge Proofs. }\label{sec:nizkp}
\ac{NIZKP} schemes allow a sender to prove a statement to a verifier, while allowing the sender to create proofs for such a statement offline, without an online interaction with the verifier.
\begin{definition}
A \ac{NIZKP} scheme $NZ$ consists of the following algorithms.\\
$NZ.Setup(1^k)$: on input a security parameter $k$, it outputs a common reference string $crs$.\\
$NZ.Proof(crs,x,w)$: on input a common reference string $crs$, a statement $x$, and a witness $w$, it outputs a proof $\pi$.\\
$NZ.Vrf(crs,x,\pi)$: on input a common reference string $crs$, a statement $x$, and a proof $\pi$, it outputs a bit $b \in {0,1}$, where $0$ indicates that the statement is not verified, while a $1$ indicates that the statement is verified.
\end{definition}
\noindent
In this paper, we use and adapt to our problem the Schnorr \ac{NIZKP} scheme~\cite{hao2017_rfc}.\\\\
\noindent
\textbf{Signatures of Knowledge. } \label{sec:sok}
\ac{SoK} schemes allow a sender, that has knowledge of a word, to sign a message while allowing a receiver to verify the knowledge of such a statement.
\begin{definition}
A \ac{SoK} scheme $SoK$ consists of the following algorithms.\\
$SoK.Setup(1^k)$: on input a security parameter $k$, it outputs a common reference string $crs$.\\
$SoK.Sign(crs,x,w,m)$: on input a common reference string $crs$, a word $x$, a witness $w$, and a message $m$, it outputs a signature $\sigma$.\\
$SoK.Vrf(crs,x,m,\sigma)$: on input a common reference string $crs$, a word $x$, a message $m$, and a signature $\sigma$, it outputs a bit $b \in {0,1}$, where $0$ indicates that the knowledge of the word $x$ is not verified, while a $1$ indicates that the knowledge of the word $x$ is verified.
\end{definition}
\noindent
In this paper, we use the SoK scheme reported in~\cite{derler2018_asiaccs} and initially defined by the authors in~\cite{faust2012_icc}.

\noindent
{\bf CPA-Full and CCA2-Full Anonymity.} In line with the current literature on anonymous group signatures, in this work, we distinguish between CPA-Full and CCA-2 Full Anonymity~\cite{derler2018_asiaccs}. 
\begin{definition}
We define a group signature scheme as \emph{CPA-Full Anonymous} if the scheme guarantees signer anonymity provided that the adversary cannot issue opening requests for specific signed messages.
\end{definition}
\begin{definition}
We define a group signature scheme as \emph{CCA2-Full Anonymous} if the scheme guarantees signer anonymity also when the adversary can issue opening requests for specific signed messages. 
\end{definition}
\noindent
It is worth noting that, without loss of generality, CCA2-Full Anonymity is stronger than CPA-Full Anonymity, as the former does not imply any constraints on the interactions between entities in the signature scheme. At the same time, as will become evident from our experimental evaluation (Sec.~\ref{sec:performance}), CCA2-Full anonymity schemes are usually more processing-intensive and energy-hungry than CPA-Full anonymity schemes, which might become relevant in constrained scenarios.

\section{Related Work}
\label{sec:related}

A few contributions investigated security and privacy issues connected with the adoption of the \ac{RID} regulation.

The authors in~\cite{svaigen2021_acm} and~\cite{svaigen2022_comcom} tried to integrate the concept of mix zones, well-known in the \ac{VANET} area for pseudonym exchange, within the UA research domain. However, mix zones require communication with infrastructure elements, which might not always be available in UA operations. The authors in~\cite{alkadi2022_tnse} proposed a decentralized traffic management protocol providing many security services, including integrity and confidentiality, and mainly focus on the security of the interactions between regulatory entities. Thus, drones' anonymity and authenticity are not considered. The authors in~\cite{hashem2021_daivna} propose the integration of the Hyperledger Iroha blockchain for the management of drones' remote identification. Based on the proposed system architecture, drones register to the blockchain using their public key and related certificate. At run-time, Internet-connected drones provide remote identification information directly on the blockchain, while drones not equipped with an Internet connection delegate the closer \ac{GCS} to write such information on the blockchain. Thus, although providing authentication and integrity of drones' messages, anonymity and privacy of drones are not considered in the design of such a solution. The authors in~\cite{brighente2022_ares} took into account the RID regulation, but focused on location privacy rather than anonymity. To this aim, they integrate the Differential Privacy (DP) tool into the \ac{RID} regulation, allowing drones to broadcast an obfuscated location, so as to preserve location privacy. Thus, the authors did not investigated specifically anonymity and message authenticity.

Besides scientific contributions, a few commercial solutions for drone remote identification are starting to appear on the market. Examples include ScaleFyt designed by Thales~\cite{scalefyt}, the Broadcast Location and Identification Platform (BLIP) designed by Unifly~\cite{blip}, and the Secure Airspace Integrated Management (SIAM) tool provided by RelmaTech~\cite{relmatech}. All these solutions rely on the authentication and anonymity services provided by the LTE cellular technology, which is, however, not the communication technology currently envisioned by the \ac{RID} regulation and the DRIP WG of the IETF.

It is worth noting that anonymity for broadcasting devices has also been considered by a few works outside the UA area. In the area of \ac{VANET}, many contributions investigated anonymity issues for vehicles, such as~\cite{calandriello2007_vanet},~\cite{benarous2020_ccnc}, and~~\cite{li2020_iotj}, just to name a few. However, such schemes either require a persistent external connection or infrastructure elements, which might not always be available for UA operations, e.g., in remote areas.

In the avionic research domain, the authors in~\cite{asari2021_comnet} proposed to generate pseudonyms for online aircrafts using an on-purpose Trusted Registration Authority, assumed to be always available online. Such an approach is not feasible for UAs, as most of them cannot rely on a persistent external connection. 
Similarly, in the maritime research domain, the authors in~\cite{goudossis2019_jmst} proposed to anonymize vessels identity using pseudonyms generated by an online trusted authority, while the authors in~\cite{hall2015_vtc} proposed to adopt the IEEE P1609.02 scheme for pseudonymous generation and disclosure. Note that standards available for pseudonyms management, such as the IEEE P1609.02, consider unicast scenarios, and not broadcast-type interactions like the ones in \ac{RID}.

Finally, we highlight that this contribution extends and improves our previous conference paper presented in~\cite{tedeschi2021_acsac} by: (i) adapting heterogeneous protocols for anonymous group signatures, traditionally adopted for online voting, to work in real-time scenarios for the anonymous identification of \acp{UA}; (ii) allowing generic receivers, namely, \emph{observers}, to verify autonomously the authenticity of the messages received by the \ac{UA}, without relying on a \ac{TTP} constantly online; (iii) reducing further the overhead of \ac{UA} anonymous identification on avionics authorities, such as the \ac{FAA}, by removing the need for expensive message authenticity management issues; and (iv) integrating our proposed \proto\ protocol suite within the standardized network architecture currently defined by the IETF WG \emph{drip}, specifically meant to ease the worldwide adoption of the \ac{RID} regulation.
Therefore, our proposed solution emerges as the first of its kind.
We will provide more details on these aspects in the next sections.

\section{Scenario and Adversary Model}
\label{sec:scenario_adv_model}

In this section, we introduce the scenario (Sec.~\ref{sec:sys_model}) and adversarial model (Sec.~\ref{sec:adv_model}) considered in our work. Moreover, we also derive the most important design and security requirements for an anonymous \ac{RID} solution (Sec.~\ref{sec:requirements}).

\subsection{System Model}
\label{sec:sys_model}

\begin{figure}[htbp]
    \centering
    \includegraphics[width=\columnwidth]{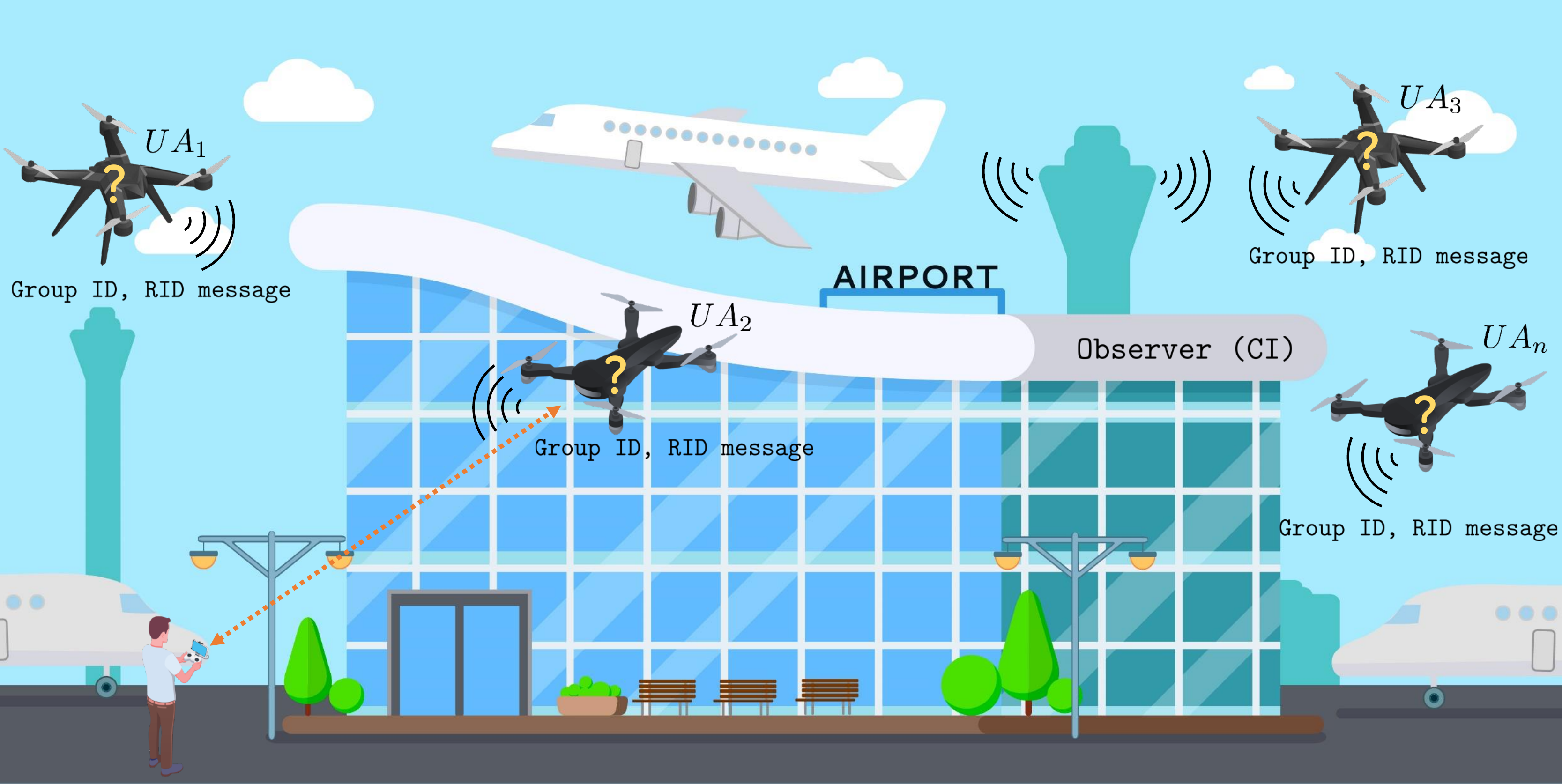}
    \caption{\color{black}{Reference scenario. Several UAs fly in a given area and broadcast messages compliant with the RID rule. Observers, such as \acp{CI}, listen to RID messages to monitor such areas and enforce protection of no-fly areas.}
    }
    \label{fig:scenario}
\end{figure}

The scenario assumed in this work, in line with the architecture defined by the \ac{DRIP} WG~\cite{drip_arch}, is depicted in Figure~\ref{fig:scenario}.

We assume several \acp{UA}, flying around a given area to complete their intended mission. The \acp{UA} are produced by different manufacturers and operated by several service providers, not trusting each other. In line with common capabilities and equipment, we assume the \acp{UA} to be equipped with a \ac{GPS} module for precise location estimation, and a wireless communication module, allowing them to emit broadcast \ac{RID} messages. Such a communication module is typically a Wi-Fi transceiver, allowing messages to be received in a radius of up to $1$~km around the \acp{UA}' location. In line with the reference architecture in~\cite{drip_arch}, we do not assume any Internet connectivity on board of \acp{UA}. This is consistent with the most general and challenging scenario, where the UA has to operate in a remote region, not offering support for Internet connectivity.

We also assume the presence of a trusted authority, namely, the \ac{USS}. In line with the definition in~\cite{drip_arch}, the USS sits in between the UAs operators and the traffic management entities, providing services such as real-time traffic monitoring and planning, data archiving, and airspace and violation control. As such, the USS hosts a \ac{PbIR}, i.e., a registry of public UA' information, to be used by observers (see definition below) to verify \ac{RID} messages emitted by \acp{UA}.

Our scenario also includes \emph{observers}, i.e., generic wireless receivers able to listen to the radio channel used by \acp{UA} to deliver wireless \ac{RID} messages. Such observers can be either standalone (i.e., smartphones of generic users and hobbyists), namely, \emph{General Public Observers}, or associated to more extensive sensor networks, such as in the case of monitoring networks of \acp{CI}, namely, \emph{Public Safety Observers}. In the latter, they monitor invasions of specific no-fly areas. They can interact online with the USS to report such invasions and allow the USS to obtain information about the identity of the UA violating flight restrictions on a given no-fly area. As an essential requirement, we also require such observers to be able to verify directly, i.e., autonomously, the authenticity of received \ac{RID} messages. Thus, although they might interact with the USS once to obtain public identification parameters of \acp{UA} stored in the \ac{PbIR}, they cannot contact the UAS regularly to verify \ac{RID} messages' authenticity. Note that this is a reasonable requirement, as having each observer interacting with the USS regularly to check every message would overload the USS, as well as require powerful resources to handle frequent cryptography operations and Internet traffic. For the sake of better readability, from now on we will always refer to \emph{Public Safety Observers} as simply observers.

Finally, for the readers' convenience, we report the main notation used below in Tab.~\ref{tab:notation}, along with a brief description.

\begin{table}[!t]
    \centering
    \caption{Notation and brief description.
    }
    \label{tab:notation}
    \begin{tabular}{P{1cm}|P{6.4cm}}
         {\bf Notation} & {\bf Description}  \\ \hline
         $UA_i$ & Generic \acl{UA}. \\
         $rcv$ & Generic observer. \\
         $T$ & \acl{USS}. \\
         $H$ & Hashing function. \\
         $p$ & Groups order. \\
         $\mathcal{G}_1$, $\mathcal{G}_2$ & Bilinear groups, with generators $P$ and $\hat{P}$. \\
         $\mathcal{G}_T$ & Prime number of $k$ bits. \\ 
         $e$ & Pairing operation. \\
         $sk_R$, $pk_R$ & Private and public key of the structure-preserving signatures on equivalence classes scheme in $DS-A^{2}RID$.  \\
         $sk_O$, $pk_O$ & Private and public key of the PKE scheme in $DS-A^{2}RID$.\\
         $sk_i$, $pk_i$ & Private and public key of the DSig scheme for the i-th UA in $DS-A^{2}RID$.\\
         $crs_j$, $crs_o$ & Reference strings of the NIZKP scheme in $DS-A^{2}RID$. \\
         $crs_S$ & Reference string of the SoK scheme in $DS-A^{2}RID$. \\
         $gpk$ & Group Public Key in $DS-A^{2}RID$. \\
         $ik$ & Issuing key in $DS-A^{2}RID$. \\
         $ok$ & Opening key in $DS-A^{2}RID$. \\
         $q$ ,$r$, $\chi$, $y$, $\rho_k$ & Nonces extracted in $\mathcal{Z}_p$. \\
         $\hat{C}_{j_i}$ & Encrypted witness in $DS-A^{2}RID$. \\
         $\sigma_{j_i}$ & Signed witness in $DS-A^{2}RID$.\\
         $reg_i$ & Registration Table entry for $UA_i$. \\
         $\sigma'$ & Signed UA credential in $DS-A^{2}RID$. \\
         $gsk_i$ & Group secret key for $UA_i$ in $DS-A^{2}RID$. \\
         $t_k$ & Timestamp of a \ac{RID} message. \\
         $pos_k$ & Location of $UA_i$ at time $t_k$. \\
         $v_k$ & Speed of $UA_i$ at time $t_k$. \\
         $em_k$ & Emergency status of $UA_i$ at time $t_k$. \\
         $\sigma_{1,k}$, $\sigma_{2,k}$ & Signatures deliverd by $UA_i$ at time $t_k$ in $DS-A^{2}RID$. \\
         $ID_g$ & Group Identifier.  \\
         & \\
         & \\
    \end{tabular}
    
\end{table}

\subsection{Adversary Model}
\label{sec:adv_model}
The adversary assumed in this contribution ($\adv$) has two main objectives, i.e., obtaining the long-term identity of a specific \ac{UA}, and impersonating a specific \ac{UA} while carrying out malicious activities using its (spoofed) identity. 

To reach such objectives, $\adv$ can carry out both passive and active attacks. On the one hand, $\adv$ is a global, frequency-unbounded, and spatially-unlimited passive eavesdropper, capable of detecting and receiving any message broadcasted by the \ac{UA}, independently from the communication frequency and modulation. We also assume $\adv$ can receive such wireless messages independently from the location of the \ac{UA}, i.e., through a network of receiving antennas deployed on the ground. On the other hand, $\adv$ can also replay previously received packets and generate ad-hoc rogue messages, trying to impersonate legitimate \acp{UA}. As such, $\adv$ possesses all the capabilities to be in line with the so-called \emph{Dolev-Yao} attacker model, commonly assumed among the most powerful attacker models considered in the literature~\cite{dolev1983}.

Note that we do not make any distinction between internal and external adversaries. Indeed, as shown in Sec.~\ref{sec:security}, being a member of a group does not provide any advantage to the adversary, and neither allows $\adv$ to successfully carry out the attacks previously mentioned.

In line with the main objectives of the \ac{RID} specification, our contribution does not take into account \ac{DoS} attacks. Indeed, the \ac{RID} architecture has been designed specifically to make spoofing and replay attacks very hard, while not dealing with \ac{DoS} attacks. In this context, our work aims to strengthen protection against replays and spoofing, while also providing anonymity to emitting \acp{UA}. 
Finally, note that our solution cannot unconditionally protect an \ac{UA} against passive tracking. Indeed, as acknowledged by several contributions in the literature, broadcasted locations can be used to link messages among them, and track a specific \acp{UA} even without being aware of its long-term identity. We refer the readers to specific works protecting locations disclosed in \ac{RID} messages for solutions to the cited specific threat~\cite{brighente2022_ares}.

\subsection{Design and Security Requirements}
\label{sec:requirements}

Based on the scenario and adversarial model described above, we can derive the following system and security requirements for a secure and anonymous \ac{RID}-compliant solution.
\begin{itemize}
    \item {\bf UA Anonymity.} To avoid privacy issues, the UA should stay anonymous while emitting \ac{RID} messages. Such a requirement implies that the long-term identity of a specific UA should never be available to observers, and observers should not be able to detect if two \ac{RID} messages are emitted by the same UA. As the observers do not need UA identification information, both CPA-Full and CCA2-Full anonymity might fulfill this requirement.
    \item {\bf UA Messages Authenticity.} All messages emitted by legitimate \acp{UA} should be verifiable as authentic, i.e., emitted by a legitimate entity and not forged. Although replay attacks are always possible due to the broadcast nature of \ac{RID} communications, they should be mitigated as much as possible. The fulfillment of this requirement is necessary to avoid false invasion claims and attributions (i.e., attempts to falsely locate a drone where it should not be allowed to fly), aimed at unveiling the identity of a specific UA.
    \item {\bf Direct Message Verification on Observers.} Observers receiving \ac{RID} messages should be able to verify the authenticity of such messages directly and autonomously, without interacting with third-parties. Such a requirement is necessary to consider observers with scarce or unstable Internet connections, as well as to avoid performance bottlenecks and excessive management on the USS.
    \item {\bf UA Identity Disclosure upon Invasion Detection.} Upon invasion detection by an observer, the USS should always be able to: (i) verify the invasion, and (ii) disclose the long-term identity of the UA causing the invasion, for further action.
\end{itemize}

In the following sections, we describe several solutions fulfilling all the above-described requirements, characterized by different features and trade-offs.

\section{The \proto\ Protocol Suite}
\label{sec:proto}

In this section, we provide the details of $A^{2}RID$, our novel protocol suite for anonymous direct authentication and remote identification of commercial \acp{UA}. We first introduce the involved actors in Sec.~\ref{sec:actors}. We then describe the $CS-A^{2}RID$ protocol in Sec.~\ref{sec:cs-scheme}, and the two variants of the $DS-A^{2}RID$ protocol in Sec.~\ref{sec:ds-scheme}.

\subsection{Actors}
\label{sec:actors}

\proto\ involves the following actors.
\begin{itemize}
    \item {\bf \ac{UA}}. It is a generic drone, following the \ac{RID} specification. As such, it emits periodic broadcast wireless messages reporting, among other information,  its location and identification. The scope of our work is to provide anonymity to the emitted messages, such that message authenticity can be verified while not revealing the long-term identity of the drone.
    \item {\bf \ac{USS} $T$.} It is an authority, in charge of releasing cryptography parameters to the UA for anonymous remote identification. The manufacturer of the UA interacts with $T$ only offline, in the Setup phase. At the same time, $T$ is available online on a public interface for queries by observers deployed by CI operators, to verify invasions of no-fly areas and unveil the location of the misbehaving UAs.
    \item {\bf Observer $rcv$.} It is a generic receiver, deployed to monitor invasions of no-fly areas of a \ac{CI}. It can directly verify the authenticity of broadcast messages emitted by \ac{UA}, without being aware of the long-term identity of the emitting entity. In case an invasion of the related no-fly area is detected, the observer can forward the received messages to the USS, for identity disclosure and enforcement of sanctions.
\end{itemize}

In the following, we describe two variants of the \proto protocol, namely, $CS-A^{2}RID$ and $DS-A^{2}RID$, suitable for integration in medium-end and low-end \acp{UA}, respectively.

\subsection{Option \#1: Camenisch-Lysyanskaya (CS) Scheme}
\label{sec:cs-scheme}
The $CS-A^2RID$ scheme, inspired by the protocol in~\cite{manulis2012group}, is a dynamic scheme with distributed authorities. The security parameter $\kappa \in \mathbb{N}$ uses cyclic groups $\mathbb{G}=\langle g \rangle$ and $\mathbb{G}_T=\langle g_T \rangle$ of prime order $\mathcal{Q}$ with $|\mathcal{Q}|=\kappa$, a bilinear map $\hat{e}:\mathbb{G}\times\mathbb{G} \rightarrow \mathbb{G}_T$, and the generator $g_T=\hat{e}(g,g)$. Moreover, it also relies on a collision-resistant hashing function $\mathcal{H}:{0,1}^*\rightarrow \mathbb{Z}_Q$. We identify four phases, i.e, the \emph{Setup Phase}, \emph{UA Joining Phase}, \emph{Online Phase}, and \emph{Disclosure Phase}.

\noindent
{\bf Setup Phase.} This phase is executed at the boot of the overall system, to setup the necessary cryptography parameters on the relevant entities, i.e., the USS and the UA. Specifically, this phase includes the following operations.
\begin{enumerate}
    \item The USS selects a security parameter $\kappa$, $x, y \in_R \mathbb{Z}_Q$, and sets $X=g^x, Y=g_T^y$.
    \item The USS also selects $h\in_R \mathbb{G}_T {\char`\\} \{1_{\mathbb{G}_T}\}, x_1,\dots, x_5 \in_R \mathbb{Z}_Q$ and sets $\mathbf{y}$ as in Eq.~\ref{eq:setupparam}.
    \begin{equation}
    \label{eq:setupparam}
        \begin{cases}
        y_1=g_T^{x_1} h^{x_2}, \\
        y_2=g_T^{x_3} h^{x_4}, \\
        y_3=g_T^{x_5}.
        \end{cases}
    \end{equation}
    Then, the USS outputs $(gpk, ik, ok, reg)$ such that:
    \begin{itemize}
        \item the group public key is set to $gpk = (\mathcal{Q}, \mathbb{G}, \mathbb{G}_T, g, g_T, \hat{e}, X, Y, h, y_1, y_2, y_3)$
        \item the secret issuing key is set to $ik = (gpk, x, y)$
        \item the secret opening key is set to $ok = (gpk, x_1, \dots, x_5)$
        \item the registration list $reg$ is initially empty.
    \end{itemize}
    \item Finally, the USS publishes the group public key $gpk$ online, to make it available to all \acp{UA} interesting in joining the system and to observers aiming to verify anonymous \ac{RID} broadcasts.
\end{enumerate}
As the USS is trusted, we assume that this phase is performed in a trusted way. Thus, the elements $x, y, h, x_1,\dots, x_5$ are chosen independently at random from $\mathbb{Z}_Q$, $\mathbb{G}_T {\char`\\}, \{1_{\mathbb{G}_T}\}$, respectively. Note that the tuple $(h, y_1, y_2, y_3)$ is a public key of the \textit{Cramer-Shoup} encryption scheme over the group $\mathbb{G}_T$ and $(x_1,\dots, x_5)$ is the corresponding private key~\cite{cramershoup_iacr}.

\noindent
{\bf UA Joining Phase.} This phase is executed offline, every time a new UA $UA_i$ would like to join the system, i.e., to start operating anonymously while complying with the RemoteID specification. This phase includes the following operations.
\begin{enumerate}
    \item $UA_i$ picks a nonce $k_i\in_R \mathbb{Z}_Q$, and it sets $P_{i,1}=g^{k_i}$. The $UA_i$ also picks a nonce $rk\in_R \mathbb{Z}_Q$, sets $R = g^{rk}$, and it computes the hash $\eta_1 = \mathcal{H}(g, R)$. Further, it sets $Sk = \eta_1 k_i + rk$, and sends the tuple $\eta_1, Sk, P_{i,1}$ via a secure channel to the USS.
    \item The USS $T$ executes the following procedures: (i) it proceeds to verify if $P_{i,1}\in\mathbb{G}$ is a point on the curve, and (ii) it computes the value $\gamma = \frac{g^{Sk}}{P_{i,1}^{\eta_1}}$ and the hash $\eta_2 = \mathcal{H}(g, \gamma)$. If $\eta_1 \stackrel{?}{=} \eta_2$ holds, it picks a nonce $r\in_R \mathbb{Z}_Q$ and computes $a_i = g^r$, $b_i = a_i^y$, and $c_i=a_i^x P_{i,1}^{rxy}$. Then, it delivers to $UA_i$ the membership certificate $(a_i, b_i, c_i)$ via a secure channel.
    \item Finally, the USS $T$ computes $P_{i,2}=\hat{e}(P_{i,1}, g)$ and stores an entry for $UA_i$ in the registration list as $reg_i=(P_{i,1}, P_{i,2})$.
    \item At message reception from the USS, $UA_i$ stores its secret signing key as $gsk_i=(gpk, k_i, a_i, b_i, c_i)$.
\end{enumerate}
Note that the tuple $(a_i, b_i, c_i)$, part of the secret signing key $gsk_i$, represents an ordinary Camenisch-Lysyanskaya signature on the element $k_i$. If all the verification steps described above end successfully, $UA_i$ stores the pair $(gsk_i,gpk)$ as its own private and public group keys, to be used for anonymously signing messages when deployed. 

\noindent
{\bf Online Phase.} This phase is executed at runtime, both on any generic observer $rcv$ and operational UA $UA_i$, which would like to send \ac{RID} messages, while staying anonymous. It involves the following operations. 
\begin{enumerate}
    \item Assume that at the time $t_k$ the UA $UA_i$ is required to deliver a \emph{RemoteID} message. We denote with $ID_g$ the UA group ID, $pos_k = (lat_k, lon_k, alt_k)$ the latitude, longitude, and altitude coordinates of the location occupied by $UA_i$, as obtained through the integrated GPS module, $pos_{u,k} = (lat_{u,k}, lon_{u,k}, alt_{u,k})$ the latitude, longitude, and altitude coordinates of the location occupied by UA Ground Station (or controller) at the time $t_k$, $v_k =(v_{x,k}, v_{y,k}, v_{z,k})$ as the 3-D vector of the speed in $m/s$ of the UA as obtained through integrated modules at the time $t_k$, and $em_k$ the emergency status value of the UA. We denote $m_k = (ID_g, pos_k, v_k, pos_{u,k}, t_k, em_k)$. 
    \item The signature generation algorithm takes as input the secret signing key $gsk_i$ of $UA_i$, the group public key $gpk$, and a generic message $m\in\{0, 1\}^*$. First, $UA_i$ computes $P_{i,2}=g_T^{k_i}=\hat{e}(P_{i,1}, g)$. Then, $UA_i$ encrypts $P_{i,2}$ under the group opener’s public key, i.e., it chooses $u\in_R \mathbb{Z}_Q$ and computes the elements of the vector $\mathbf{T}$ as in Eq.~\ref{eq:signparam0}.
    \begin{equation}
    \label{eq:signparam0}
        \begin{cases}
        T_1 = g_T^u, \\
        T_2 = h^u, \\
        T_3 = y_1^u P_{i,2}, \\
        T_4 = y_2^u y_3^{u\mathcal{H}(T_1||T_2||T_3)}.
        \end{cases}
    \end{equation}
    
    \item Then, $UA_i$ selects $r,r'\in_R \mathbb{Z}_Q$ and computes a blinded version of the certiﬁcate, namely, $\tilde{\sigma}=(T_5, T_6, T_7)$, according to Eq.~\ref{eq:signparam1}.
    \begin{equation}
    \label{eq:signparam1}
        \begin{cases}
        T_5=a_i^{r'}, \\
        T_6=b_i^{r'}, \\
        T_7=c_i^{r'r}.
        \end{cases}
    \end{equation}
    \item $UA_i$ also picks a nonce $\rho, \mu, \nu\in_R \mathbb{Z}_Q$, and performs the operations in Eq.~\ref{eq:signparam2}.
    \begin{equation}
    \label{eq:signparam2}
        \begin{cases}
        R_1 = \frac{\hat{e}(g, T_7)^{\rho}}{\hat{e}(X, T_6)^{\mu}}, \\
        R_2 = g_T^{\nu}, \\
        R_3 = h^{\nu}, \\
        R_4 = y_1^{\nu} g_T^{\mu}, \\
        R_5 = y_3^{\nu\mathcal{H}(T_1||T_2||T_3)}.
        \end{cases}
    \end{equation}
    Then, using the plaintext \ac{RID} message $m_k$, the $UA_i$ computes the hash of the generated parameters and the message $m_k$ as in Eq.~\ref{eq:signparam3}.
    \begin{equation}
    \label{eq:signparam3}
        \begin{cases}
        R = (R_1,R_2,R_3,R_4,R_5), \\
        c = \mathcal{H}(R,g,g_T,X,Y,h,y_1,y_2,y_3,m_k), \\
        S_{\rho}=\frac{c}{r} + \rho, \\
        S_{\mu}=ck_i + \mu, \\
        S_{\nu}=cu + \nu.
        \end{cases}
    \end{equation}
    
    \item Finally, $UA_i$ computes the signature $\sigma_k$ over the message $m_k$ as in Eq.~\ref{eq:signature}.
    \begin{equation}
        \label{eq:signature}
        \sigma_k = (S_{\rho}, S_{\mu}, S_{\nu}, T_1, T_2, T_3, T_4, T_5, T_6, T_7).   
    \end{equation}
    
    $UA_i$ delivers in broadcast the message $m_k$, the signature $\sigma_k$, and the value $mode_k=0$, indicating the usage of the $CS-A^2RID$ protocol of the $A^2RID$ protocol suite. Note that, in the above signature generation algorithm, the values $(T_1, T_2, T_3, T_4)$ are a Cramer-Shoup encryption of $P_{i,2}$ under the group opener’s public key $(h, y_1, y_2, y_3)$~\cite{cramershoup_iacr}. The ciphertext can be decrypted using the private key $(x_1,\dots,x_5)$. Thus, the signature $S$ proves that the signer is in possession of a $k_i$, satisfying $P_{i,1} = g^{k_i}$ and proving, in turn, that the signer has a valid signing key $gsk_i$.
    \item On the receiving side, every observer $rcv$ receiving the message can directly authenticate it and verify its contents. Specifically, $rcv$ extracts the message $m_k$ and parses the signature $\sigma_k$ and the mode $mode_k$. If $mode_k=1$, indicating that the message is secured with $CS-A^2RID$, it proceeds as in the next steps.
    \item First, $rcv$ computes the hash $H=\mathcal{H}(T_1||T_2||T_3)$, and the following parameters, as in Eq.~\ref{eq:verifyparam}.
    \begin{equation}
    \label{eq:verifyparam}
        \begin{cases}
        R_1' = \frac{\hat{e}(g, T_7)^{S_{\rho}}}{\hat{e}(X, T_6)^{S_{\mu}} \hat{e}(X, T_5)^H}, \\
        R_2' = \frac{g_T^{S_{\nu}}}{T_1^H}, \\
        R_3' = h^{S_{\nu}} - T_2^H, \\
        R_4' = (y_1^{S_{\nu}} g_T^{m_k}) - T_3^H, \\
        R_5' = \frac{y_2^{S_{\nu}}y_3^{HS_{\nu}}}{T_4^H}.
        \end{cases}
    \end{equation}
    \item Then, $rcv$ computes the hash of the message as in the following Eq.~\ref{eq:verifyhash}.
    \begin{equation}
    \begin{cases}
        \label{eq:verifyhash}
        R' = (R_1',R_2',R_3',R_4',R_5'), \\
        c' = \mathcal{H}(R',g,g_T,X,Y,h,y_1,y_2,y_3,m_k).
    \end{cases}
    \end{equation}
    \item If $c\stackrel{?}{=}c'$ applies and $\hat{e}(T_5, Y)\stackrel{?}{=}\hat{e}(g, T_6)$, the signature is deemed valid and authentic. Otherwise, the signature algorithm outputs an error, rejecting the message.
\end{enumerate}

\noindent
{\bf Disclosure Phase.} This phase is triggered by the generic observer $rcv$ and executed on the USS, whenever an invasion of the no-fly area of $rcv$ by an anonymous UA is detected. It involves the following operations.
\begin{enumerate}
    \item Assume $rcv$ realizes that the received message $(m, mode, \mathbf{\sigma_k})$, verified as authentic, determines an invasion of the monitored no-fly zone. Then, $rcv$ forwards this message to the USS $T$. 
    \item The USS retrieves the secret opening key $ok$ and the group public key $gpk$ corresponding to the message $m_k$, as well as the registration list $reg$. First, the USS verifies the signature $\sigma_k$, by executing the signature verification algorithm, checking that $T_4 \stackrel{?}{=} T_1^{x_3+x_5H}T_2^{x_4}$. If the check is verified, it proceeds further; otherwise, it outputs an error and rejects the received message.
    \item Then, the USS computes the value $P_{i,2} = \frac{T_3}{(T_1^{x_1} T_2^{x_2})}$, and it finds the UA $UA_i$ such that $reg_i =(P_{i,1}, P_{i,2})$. If no such $UA_i$ exists, it outputs an error. Otherwise, the USS can identify $UA_i$ and return an acknowledgment to $rcv$, without revealing the identity of such UA.
\end{enumerate}

\subsection{Option \#2: Derler-Slamanig (DS) Schemes}
\label{sec:ds-scheme}

Although the $CS-A^2RID$ scheme described in the previous section allows \acp{UA} to sign \ac{RID} messages anonymously, it requires pairing operations during signature generation, hardly achievable on very constrained \acp{UA} within the time bounds defined by the \ac{RID} specification ($1$~s). To this aim, in this section, we introduce the adaptation of the Derler-Slamanig (DS) schemes proposed in~\cite{derler2018_asiaccs} to our scenario, namely, $DS-CCA2-A^{2}RID$ and $DS-CPA-A^{2}RID$. We anticipate that, as a distinguishing feature, both $DS-CCA2-A^{2}RID$ and $DS-CPA-A^{2}RID$ do not require the signing UA to use pairing operations, thus being feasible for integration into constrained \acp{UA} for anonymous remote identification. In the following, we describe all the operations required by the protocols in our setting. We delve into the details of the instantiation of the structure-preserving signatures on equivalence classes, SoK, and NIZKP scheme while, due to their more generic features and ease of notation, we report the public-key encryption setup, public-key encryption, public key decryption, digital signature scheme setup, digital signature, and digital signature verification operations through the general notation $PKE.KGen(\cdot)$ $PKE.Enc(\cdot)$, $PKE.Dec(\cdot)$, $DSig.KGen(\cdot)$, $DSig.Sgn(\cdot)$, and $DSig.Vrf(\cdot)$, respectively. Interested readers can find the details of the used PKE and DSig schemes in~\cite{certicom2009} and~\cite{boneh2003_tact}, respectively.

\textcolor{black}{In line with the previous scheme, we identify four protocol phases, i.e, the \emph{Setup Phase}, \emph{UA Joining Phase}, \emph{Online Phase}, and \emph{Disclosure Phase}}.

\noindent
{\bf Setup Phase.} This phase is executed at the boot of the overall system, to setup the necessary cryptography parameters on the relevant entities, i.e., the USS and the UA. Specifically, this phase includes the following operations.
\begin{itemize}
    \item The USS $T$ takes a bilinear group $BG = (p, \mathcal{G}_1, \mathcal{G}_2, \mathcal{G}_T, e, P, \hat{P})$, where $p$ is the order of the groups $\mathcal{G}_1$ and $\mathcal{G}_2$, $\mathcal{G}_T$ is a prime number of bit-size $\kappa$, $e$ is a pairing, and $P$ and $\hat{P}$ are generators of $\mathcal{G}_1$ and $\mathcal{G}_2$, respectively.
    \item The USS $T$ generates the key-pair of the structure-preserving signatures on equivalence classes scheme, namely, $(sk_R, pk_R)$. To this aim, on input $BG$ and a vector length $l$, it chooses $(x_i)_{i \in |l|} \leftarrow (\mathcal{Z}_p)^l$, sets $sk_R = (x_i)_{i \in |l|}$, and computes $pk_R = (x_i\hat{P})_{i \in |l|}$.
    \item The USS $T$ initializes the key pair of the \ac{PKE} algorithm, namely, $(sk_O, pk_O)$, as $(sk_O, pk_O) \leftarrow PKE.KGen(1^k)$.
    \item The USS $T$ generates two reference strings of the \ac{NIZKP} scheme, i.e., $crs_j$ and $crs_O$, through the application of a string generation algorithm $CRSGen(gk)$, being $gk=(\mathcal{G},q,g)$, where $\mathcal{G}$ is a group of prime order $q$ and generator $g$.
    \item The USS $T$ generates the reference string of the \ac{SoK} scheme, i.e., $crs_S$, through the string generation algorithm $CRSGen(gk)$.
    \item The USS sets the group public key $gpk = (pk_R, pk_O, crs_j, crs_O, crs_S)$, the issuing key $ik = sk_R$, and the opening key $ok = sk_O$.
    \item Finally, the USS publishes the group public key $gpk$ online, to make it available to all \acp{UA} interested in joining the system and to observers aiming to verify anonymous \ac{RID} broadcasts.
\end{itemize}

At the same time, the generic UA $UA_i$ initializes the key pair of the \ac{DSig} algorithm, namely, $(sk_i, pk_i)$, as $(sk_i, pk_i) \leftarrow DSig.KGen(1^K)$. Then, it makes $pk_i$ publicly available.

\noindent
{\bf UA Joining Phase.} This phase is executed offline, every time a new UA $UA_i$ would like to join the system, i.e., to start operating anonymously while complying with the RemoteID specification. This phase includes the following operations.
\begin{itemize}
    \item Assume $UA_i$ would like to obtain the cryptography materials necessary to join the group.
    $UA_i$ first extracts two nonces $(q,r) \leftarrow \mathcal{Z}_p^*$. Then, it computes the elliptic curve points $Q = qP$ and $U_i = r \cdot qP$. Then, it generates the encrypted witness $\hat{C}_{j_i}$ as $\hat{C}_{j_i} = PKE.Enc(pk_O,r\hat{P})$, and generates the signed witness $\sigma_{j_i}$ as $\sigma_{j_i} = DSig.Sgn(sk_i,\hat{C}_{j_i})$. 
    Finally, $UA_i$ generates the proof $\pi_{j_i}$ by applying the proof algorithm of the \ac{NIZKP} scheme. To this aim, it extracts a nonce $\chi \leftarrow Z_p$, computes $A=\chi Q$, $B=\chi \hat{P}$, $c=H(Q,U,\hat{P},C_{j_i},A,B)$, and $s=\chi-c\cdot r$, and sets $\pi_{j_i} = (c,s)$.\\
    The values $M_j = ( (U_i, Q), \hat{C}_{j_i}, \sigma_{j_i}, \pi_{j_i})$ and $st = (gpk, q, U_i, Q)$ are then delivered to the authority $T$.
    \item At reception of the message, the USS first stores in the \emph{Registration Table} the entry $reg_i = (\hat{C}_{j_i}, \sigma_{j_i})$.
    \item Then, the USS $T$ generates the signature $\sigma'$ by applying the structure-preserving signatures on equivalence classes scheme. Specifically, on input the plaintext $m=((U_i, Q))$ and the secret key $sk_R=(x_i)_{i \in |l|}$, the USS extracts $y \leftarrow Z_p$ and generates the signature $\sigma=(Z,Y,\hat{Y})$, being $Z= y \sum_{i \in |l|} x_iM_i$, $Y=\frac{1}{y}P$, and $\hat{Y}=\frac{1}{y}\hat{P}$. 
    \item Then, the USS $T$ generates a new anonymous identifier for $UA_i$, by applying the algorithm $S.ChgRep$. Specifically, on input the message $m=(U_i,Q)$, the signature $\sigma'=(Z,Y,\hat{Y})$, the scalar $q^{-1}$, and the public key $pk_R$, the USS picks $\psi,\mu \leftarrow \mathcal{Z}_p$ and outputs $\sigma = (\psi \mu Z, \frac{1}{\psi}Y, \frac{1}{\psi}\hat{Y})$. The USS then constructs group secret key for $UA_i$ as $gsk_i = ((R=rP,P),\sigma)$.
    \item The signature $\sigma'$ and the group secret key $gsk_i$ are then delivered to $UA_i$.
    \item Finally, $UA_i$ verifies all the received signatures. First, it verifies the NIZKP signature $\pi_{j_i}$. To this aim, it first parses $\pi_{j_i} = (c,s)$, and then computes $\hat{A} = (sQ + cU)$ and $\hat{B} = (sP + cC_{j_i})$. Then, it checks if $c \stackrel{?}{=} H(Q,U,\hat{P},C_{j_i},\hat{A},\hat{B})$. If the check is verified, it proceeds further. \\
    Second, it verifies the digital signature $\sigma_{j_i}$. To this aim, it checks that $ DSig.Vrf(pk_i,\hat{C}_{j_i},\sigma_{j_i}) \stackrel{?}{=} 1$. \\
    Finally, it verifies the signature of the structure-preserving signature on equivalence classes scheme. To this aim, on input $m=(m_i)_{i \in l}=(U_i,Q)$, the signature $\sigma'=(Z,Y,\hat{Y})$, and the public key $pk_R$, it checks that the following Eq.~\ref{eq:vrf_r} holds.
    \begin{align} 
    \label{eq:vrf_r}
        \prod_{i \in |l|} e(m_i,\hat{X_i}) = e(Z,\hat{Y}) \nonumber \\ 
        e(Y,\hat{P}) = e(P,\hat{Y}),
    \end{align}
    being $\hat{X_i} \in \mathcal{G}_2^*$, and $e (\cdot,\cdot)$ the bilinear pairing operation. \\
\end{itemize}

If all the verification steps described above end successfully, $UA_i$ stores the pair $(gsk_i,gpk)$ as its own private and public group keys, to be used for anonymously signing messages when deployed. 

\noindent
{\bf Online Phase.} This phase is executed at runtime, both on any generic observer $rcv$ and operational UA $UA_i$, which would like to send \ac{RID} messages, while staying anonymous. It involves the following operations.
\begin{itemize}
    \item Assume that at the time $t_k$ the UA $UA_i$ is required to deliver a \emph{RemoteID} message. We denote with $ID_g$ the UA group ID, $pos_k = (lat_k, lon_k, alt_k)$ the latitude, longitude, and altitude coordinates of the location occupied by $UA_i$, as obtained through the integrated GPS module, $pos_{u,k} = (lat_{u,k}, lon_{u,k}, alt_{u,k})$ the latitude, longitude, and altitude coordinates of the location occupied by UA Ground Station (or controller) at the time $t_k$, $v_k =(v_{x,k}, v_{y,k}, v_{z,k})$ as the 3-D vector of the speed in $m/s$ of the UA as obtained through integrated modules at the time $t_k$, and $em_k$ the emergency status value of the UA. We denote $m_k = (ID_g, pos_k, v_k, pos_{u,k}, t_k, em_k)$. The following operations depend on the selected operational mode, i.e., $DS-CCA2-A^{2}RID$ or $DS-CPA-A^{2}RID$.
    \item $\bm{DS-CCA2-A^{2}RID}$. $UA_i$ extracts a nonce $\rho_k \leftarrow Z_p$, and computes two signatures for the time-slot $k$, namely, $\sigma_{1,k}$ and $\sigma_{2,k}$. \\
    $\sigma_{1,k}$ is obtained by applying the randomization algorithm of the structure-preserving signatures on equivalence classes scheme on the group secret key $gsk_i$, i.e., it sets $\sigma_k = \sigma = (Z,Y,\hat{Y})$, extracts $\phi_k \leftarrow Z_p$, and computes $\sigma_{1,k} = (\phi_k \mu Z, \frac{1}{\phi_k}Y, \frac{1}{\phi_k}\hat{Y}) = ( (R',P'),\sigma')$.\\
    $\sigma_{2,k}$ is obtained by signing the message $m_k$ through the signature algorithm of the SoK scheme, Specifically, $UA_i$ extracts $u_k,v_k,\eta_k \leftarrow Z_p$, computes $(\hat{C_1},\hat{C_2}) = (u_k\hat{Y},\rho\hat{P} + u\hat{P})$, $N = v_kP$, $\hat{M}_1=\eta_k\hat{Y}$,$\hat{M}_2=(v_k + \eta_k)\hat{P}$, $c=H(N||\hat{M}_1||\hat{M}_2||\sigma_{1,k}||m)$, $z_1=v_k+c\cdot \rho_k$, $z_2=\eta+c\cdot u_k$, and finally, it sets $\sigma_{2,k}=(\hat{C}_1, \hat{C}_2, c, z_1, z_2)$.\\
    Finally, $UA_i$ delivers the RemoteID message $(m_k, mode=1, \mathbf{\sigma_k}=(\sigma_{1,k}, \sigma_{2,k}))$.
    \item $\bm{DS-CPA-A^{2}RID}$. $UA_i$ extracts two nonces $\rho_k,v_k \leftarrow Z_p$, and computes two signatures for the time-slot $k$, namely, $\sigma_{1,k}$ and $\sigma_{2,k}$. \\
    $\sigma_{1,k}$ is obtained as for the $DS-CCA2-A^{2}RID$ scheme above. $\sigma_{2,k}$ is the pair $(c_k,z_k)$, where $c_k = H( N||\sigma_{1,k})||m_k)$, with $N=v_k P$, and $z_k = v_k + c_k \cdot \rho_k$. Finally, $UA_i$ delivers the RemoteID message $(m_k, mode=2, \mathbf{\sigma_k}=(\sigma_{1,k}, \sigma_{2,k}))$.
    \item Assume the generic observer $rcv$ receives the message $(m_k, mode, \mathbf{\sigma_{k}}$, and would like to verify its authenticity.
    \item If the group public key corresponding to the group identifier $ID_g$ is not stored locally, $rcv$ retrieves it online from the USS $T$. If $gpk$ has already been retrieved, this step is not necessary.
    \item If $mode=1$, the message was broadcasted in CCA2 mode. Thus, $rcv$ verifies the signature components $(\sigma_{1,k}, \sigma_{2,k})$ by applying individually the verification algorithms of the structure-preserving signatures on equivalence classes and SoK schemes on $\sigma_{1,k}$ and $\sigma_{2,k}$, respectively. \\
    First, $rcv$ checks the authenticity of $\sigma_{1,k}$. To this aim, it sets $\sigma_{1.k}={(R',P'),\sigma} = (Z,Y,\hat{Y})$ and $pk_R=(\hat{X}_i)_{i \in |l|}$, and checks that Eq.~\ref{eq:vrf_r} reported above holds. In case it does not hold, the RemoteID message is not verified, and it is reported as anomalous.\\
    If Eq.~\ref{eq:vrf_r} holds, then $rcv$ sets $N' = z_1P - cP'$, $M_1 = z_2 \cdot \hat{Y} - c \cdot \hat{C}_1$, and $\hat{M}_2=(z_1+z_2)\cdot \hat{P} - c\cdot \hat{C}_2$, and checks if $c \stackrel{?}{=} H(N'||\hat{M}_1||\hat{M}_2||\sigma_{1,k})||m_k)$. If it holds, the message $m$ is authenticated. Otherwise, it is discarded as not authentic.
    \item If $mode=2$, the message was broadcasted in CPA mode. As previously-described, $rcv$ verifies the signature components $(\sigma_{1,k}, \sigma_{2,k})$ individually. Thus, $rcv$ checks the authenticity of $\sigma_{1,k}$, using the same procedure described above for the $DS-CCA2-A^{2}RID$. If Eq.~\ref{eq:vrf_r} does not hold, it sets $N' = zP - cP'$, and checks if $c \stackrel{?}{=} H(N'||\sigma_{1,k}||m_k)$. If the equality holds, the message is authenticated, otherwise, the message is not authenticated and, it is discarded.
\end{itemize}

\noindent
{\bf Disclosure Phase.} This phase is triggered by the generic observer $rcv$ and executed on the USS, whenever an invasion of the no-fly area of $rcv$ by an anonymous UA is detected. It involves the following operations.
\begin{itemize}
    \item Assume $rcv$ realizes that the received message $(m, mode, \mathbf{\sigma_k})$, verified as authentic, determines an invasion of the monitored no-fly zone. Then, $rcv$ forwards this message to the USS $T$. 
    \item On message reception, $T$ finds the entry $reg_i \leftarrow (\hat{C}_{j_i}, \sigma_{j_i})$ such that the following Eq.~\ref{eq:open} applies.
    \begin{align}
        \label{eq:open}
        \hat{R} = PKE.Dec(ok,\hat{C}_{j,i}) \nonumber \\
        e(\sigma_{1,k}[1][1],\hat{P}) = e(\sigma_{1,k}[1][2],\hat{R}).
    \end{align}
    \item $T$ finally returns an ack message to the observer, to confirm the correct execution of identification operations. Recall that $T$ does not reveal the identity of the invading $UA$, as $rcv$ is not the entity in charge of enforcing sanctions, charges, or bans.
\end{itemize}

\section{Security Considerations}
\label{sec:security}
\textcolor{black}{In this section, we discuss the security features offered by the protocols in \proto. Sec.~\ref{sec:sec_obj} provides the big picture of the security services offered by the \proto\ protocol suite, while Sec.~\ref{sec:formal} reports the details of their formal security verification conducted using \pcalgostyle{ProVerif}.}

\subsection{Security Services}
\label{sec:sec_obj}

Overall, the protocols in \proto\ provide the following security properties.\\
\noindent
\textbf{UA Anonymity.} One of the most important security properties provided by the protocols in the \proto\ protocol suite is the sender's full anonymity in the Random Oracle Model under the \ac{DDH} assumption in $\mathbb{G}_T$. In the $CS-A^2RID$ scheme, even assuming that the adversary knows the secret signing keys $gsk_i$ of all group members, he/she cannot distinguish if the signature $\sigma = (S_{\rho}, S_{\mu}, S_{\nu}, T_1, T_2, T_3, T_4, T_5, T_6, T_7)$ has been generated by two different signers. Indeed, the corresponding signature features the statistical Zero Knowledge property, i.e., it does not reveal any information about the secret signing key. Thus it is computationally hard (under the \ac{DDH} assumption) for the adversary to distinguish which group secret key has been used to compute the elements $T_1, T_2, T_3, T_4$. Moreover, since the values $T_5, T_6, T_7$ are blinded, the adversary cannot extract any information about the membership certificate of the user.
Similarly, in the two variants of the $DS-A^2RID$ scheme, the two signature components $\sigma{1,k}$ and $\sigma{2,k}$ are obtained by using the group public key and a pseudo-random transformation of the group secret key $gsk_i$ through the structure-preserving signatures on equivalence classes algorithm. By definition, two elements of the same equivalent class are unlinkable, i.e., it is not possible to attribute any of these elements to a single (or multiple) entity(ies), thus providing sender anonymity.
In Sec.~\ref{sec:formal}, we formally verify the anonymity of the sender for all the above protocols using \pcalgostyle{ProVerif}.

\noindent
\textbf{UA Message Authenticity.} Another important security feature offered by \proto\ is the capability of generic observers to authenticate directly the received anonymous broadcast messages, without relying on any \acp{TTP}. In the $CS-A^2RID$ scheme, the observer uses the group public key $gpk$ to verify that a legitimate signer, in possession of the group secret key $gsk_i$, signed the message $m$, thus proving the authenticity of the message. 
Similarly, in the two variants of the $DS-A^2RID$ scheme, the observers can use the group public key $gpk$ to ensure that a legitimate correspondent group secret key $gsk_i$ has been used to sign the message, thus guaranteeing message authenticity. This is a significant security feature, especially for deployability purposes, as it allows observers not to be Internet-connected and to detect misbehaving \acp{UA} immediately. We also formally verify the authenticity of received messages for all the above protocols using \pcalgostyle{ProVerif} in Sec.~\ref{sec:formal}.

\noindent
\textbf{Protection against Replay Attacks.} As $A^2RID$ messages are broadcast, replay attacks cannot be avoided, but only detected and mitigated through careful protocol integration and deployment. In \proto\ protocols, protection against replay attacks is provided thanks to the timestamp $t_s$ included in regular standard-compliant \ac{RID} messages. At message reception time, the observer decodes the timestamp and, if the difference between the information in the message and the actual reception time is higher than a threshold $\tau$, it discards the message. At the same time, thanks to the message authenticity feature discussed above, any malicious modification of the timestamp in the message leads to the failure of the verification of the message signature(s), leading to message discarding.

\noindent
\textbf{Partial Protection against UA tracking.} Although $A^2RID$ specifically addresses sender anonymity, in our context, it also provides partial protection against passive tracking of the \ac{UA}. Indeed, in the standard setup where the eavesdropper does not know how many \acp{UA} are flying in the area corresponding to its reception range, it is not possible to associate two (or more) messages to the same \ac{UA} with $100$\% assurance. \\
However, as highlighted also by the authors in~\cite{sun2017_cns} and ~\cite{khakpour2017_vcomm}, working on disclosed locations, an eavesdropper with capabilities to run data analysis might be able to infer the track of a given \ac{UA} with remarkable accuracy, resorting to state-of-the-art Artificial Intelligence (AI) techniques. We acknowledge the likely effectiveness of these solutions for \ac{UA} tracking and thus, we define our protection against tracking as \emph{partial}. However, we recall that the aim of the $A^2RID$ protocol suite is to provide \acp{UA} anonymity and messages authenticity for \acp{UA} compliant with the \ac{RID} specification. Therefore, location privacy is out of scope and can be provided through different solutions, easily integrable in \ac{RID} messages~\cite{brighente2022_ares}.

\subsection{Formal Verification using \pcalgostyle{ProVerif}}
\label{sec:formal}

\textcolor{black}{We formally verified the security features offered by \proto, i.e., \ac{UA} anonymity and \ac{RID} message authenticity, through the tool \pcalgostyle{ProVerif}~\cite{proverif}. Note that the rigorous security proofs of the CS and DS schemes adopted in our manuscript have been provided by the authors in~\cite{manulis2012group} and~\cite{derler2018_asiaccs}, respectively. However, the application of such schemes in the drone ecosystem and their combination with additional security properties might, in principle, \emph{logically} affect the security of such schemes. In this context, using automated security protocol verification tools such as ProVerif is the correct way to check if the way such protocols are used does not affect their security properties. \\
In brief, \pcalgostyle{ProVerif} is an automatic cryptographic protocol verifier widely adopted in the literature to formally verify the security properties achieved by cryptographic protocols~\cite{tedeschi2022_tdsc_ppca},~\cite{wu2022_sp},~\cite{zhang2022_tifs}. 
}

Specifically, \pcalgostyle{ProVerif} assumes the Dolev-Yao attacker model, i.e., the attacker can read, modify, delete, and forge new packets to be delivered on the communication channel. Assuming the aforementioned conditions, \pcalgostyle{ProVerif} checks if the attacker can break the security goals of the protocol, as defined by the user. In case an attack is found, \pcalgostyle{ProVerif} also provides a step-by-step description of the attack. \textcolor{black}{Internally, ProVerif uses an abstract representation of the protocols through Horn clauses, and it adopts a logical resolution algorithm on these clauses to prove the security properties of the protocol or to find attacks. ProVerif also translates the security properties of interest into \emph{derivability queries} on the generated Horn clauses, and it finds attacks (if any) working on the reachability of such clauses. Interested readers can find all the details on the internal working logic of ProVerif in the seminal work in~\cite{blanchet2022_eptcs}.
}

We formally verified \proto\ protocols in \pcalgostyle{ProVerif} to prove: (i) the secrecy of the long-term identity of the \ac{UA}; and, (ii) the authenticity of the message broadcasted by an \ac{UA}. Therefore, according to the logic of the \pcalgostyle{ProVerif} tool, we defined two main events.
\begin{enumerate}
    \item \emph{acceptUA(id)}, indicating that an UA with long-term identity $ID_n$ is running a protocol in \proto.
    \item \emph{termAuth(id)}, indicating that the USS finished successfully executing a protocol in \proto, verifying that the UA $ID_n$ was the one to generate the message.
\end{enumerate}
Moreover, we proved the message authenticity property by verifying security properties such as \emph{sender authentication} and \emph{impersonation resistance}. As standard mechanisms in \pcalgostyle{ProVerif} we checked that \emph{event(acceptUA(id))} cannot be executed after the execution of \emph{event(termAuth(id))}.

Furthermore, we verified the strong secrecy of the long-term identity of $UA_i$, ensuring that the attacker cannot to distinguish when the secret changes, and that the attacker cannot obtain $UA_i$ from the messages exchanged on the wireless communication channel.
The following output messages are provided by \pcalgostyle{ProVerif} to identify the fulfillment of the security properties of our interest.
\begin{itemize}
    \item \emph{event(last\_event ()) $\Longrightarrow$ event(previous\_event ()) is true}: meaning that the function \emph{last\_event} is executed only when another function, namely \emph{previous\_event}, is really executed;
    \item \emph{not attacker(elem[]) is true}: meaning that the attacker is not in possession of the value of \emph{elem};
    \item \emph{Non-interference elem[] is true}: meaning that an attacker cannot infer the value of \emph{elem} from the eavesdropped messages.
\end{itemize}
We anticipate that the interested readers can verify our claims and reproduce our results by downloading the source code of the security verification in \pcalgostyle{ProVerif} of both the $CS-A^2RID$ protocol and the $DS-A^2RID$ protocols at~\cite{arid2_code} and~\cite{arid2_code_Eva}, respectively. 

\noindent
{\bf $\bm{CS-A^{2}RID}$ Scheme.} We first implemented $CS-A^{2}RID$ in \pcalgostyle{ProVerif} and, we tested the aforementioned two security properties. An excerpt of the output of the \pcalgostyle{ProVerif} tool for the case of $CS-A^{2}RID$ is shown in Figure~\ref{fig:proverif_cs_arid}.
\begin{figure}[htbp]
    \centering
    \begin{tcolorbox}[title=Verification Summary - $CS-A^2RID$]
        \texttt{Query \pcalgostyle{event(termAuth(id))} ==> \pcalgostyle{event(acceptUA(id))} is \textbf{true}.}\\
        \texttt{Query \pcalgostyle{Query not attacker(IDA[])} is \textbf{true}.}\\
        \texttt{Non-interference IDA is \textbf{true}}.
    \end{tcolorbox}
    \caption{Excerpt of the output provided by the \pcalgostyle{ProVerif} tool for verifying the security properties of the $CS-A^2RID$ protocol.}
    \label{fig:proverif_cs_arid}
\end{figure}
The tool confirms that: (i) message authenticity is verified; (ii) the attacker cannot obtain the long-term identity of the UA, namely, $UA_i$; and, (iii) the attacker cannot even distinguish when the emitter of a given message changes. Thus, $CS-A^{2}RID$ achieves the security properties of our interest.

\noindent\\
{\bf $\bm{DS-A^{2}RID}$ Schemes.} We also implemented both the $DS-A^{2}RID$ schemes in \pcalgostyle{ProVerif}, and we tested also for these protocols the same security properties verified for the $CS-A^{2}RID$ scheme. We report in Fig.~\ref{fig:proverif_ds_cca2_arid} the excerpt of the output of the \pcalgostyle{ProVerif} tool for both the $DS-CCA2-A^{2}RID$ and the $DS-CPA-A^{2}RID$ protocols.
\begin{figure}[htbp]
    \centering
    \begin{tcolorbox}[title=Verification Summary - $DS-A^2RID$]
        \texttt{Query \pcalgostyle{event(sig\_verified(id))} ==> \pcalgostyle{send\_message(id))} is \textbf{true}.}\\
        \texttt{Query \pcalgostyle{Query not attacker(secret\_gsk[])} is \textbf{true}.}\\
        \texttt{Non-interference secret\_gsk is \textbf{true}}.
    \end{tcolorbox}
    \caption{Excerpt of the output provided by the \pcalgostyle{ProVerif} tool for verifying the security properties of the $DS-CCA2-A^2RID$ protocol.}
    \label{fig:proverif_ds_cca2_arid}
\end{figure}

We first notice that whenever a message by the UA $id$ is authenticated by an observer (\pcalgostyle{event(sig\_verified(id)}), that message has been actually delivered by the UA with unique identifier $id$, thus verifying message authenticity. Such authenticity verification always occurs directly, without relying on trusted third parties. Moreover, we also notice that the group secret key $secret\_gsk$ used by the UA to anonymize its messages is never leaked to the adversary (\pcalgostyle{Query not attacker(secret\_gsk[])} is \textit{true}), and that such an adversary cannot even discriminate if the key used to anonymize two or more messages is the same or not (Non-interference $secret\_gsk$ is \textit{true}). Thus, the main security properties to be provided by $DS-CCA2-A^2RID$ have been verified. Note that the same properties have also been verified for the $DS-CPA-A^2RID$ scheme. The interested readers can verify our claims and reproduce our results by downloading the source code of the security verification in \pcalgostyle{ProVerif} at the link provided in~\cite{arid2_code_Eva}.

\section{Performance Assessment}
\label{sec:performance}

In this section, we present the results of our extensive experimental performance assessment, performed on actual \acp{UA}. Sec.~\ref{sec:impl} provides the implementation details, while Sec.~\ref{sec:results} presents all the experimental results.

\subsection{Implementation Details}
\label{sec:impl}

{\bf CS Scheme.} We implemented a prototype of the $CS-A^2RID$ protocol on the Holybro X-500 commercial drone~\cite{holybro_x500}. The Holybro X-500 features a mission computing unit UP Xtreme i7 8665UE, connected to a \emph{Pixhawk 4 Autopilot}. The \emph{Pixhawk 4} is connected, in turn, to a Holybro M8N GPS module. It also includes a quad-core processor $1.70$~GHz Intel Core i7, $64$~GB of eMMC memory, and $16,384$~MB of RAM, being definitely a high-end \ac{UA}. As for the \ac{OS}, the drone runs \emph{Ubuntu 20.04 LTS Focal Fossa}~\cite{ubuntufossa}. \\
We implemented $CS-A^2RID$ in \emph{C++}, and we integrated it at the IEEE 802.11b MAC-layer. 
In particular, we integrated \proto\ within a custom IEEE 802.11b MAC-layer by using the network packet sniffing and crafting library libtins 4.0~\cite{fontanini2016libtins} and the lightweight Micro Air Vehicle Message Marshalling Library MAVSDK 1.4.3~\cite{mavsdk}, used to acquire the GPS coordinates and drone speed. We also recall that libtins allows full customization of IEEE 802.11 raw MAC-layer messages, such as broadcast data frames encapsulated within IEEE 802.11 standard with a Maximum Transmission Unit (MTU) of $2,312$~B. 

\begin{figure}[htbp]
    \centering
    \includegraphics[width=\columnwidth]{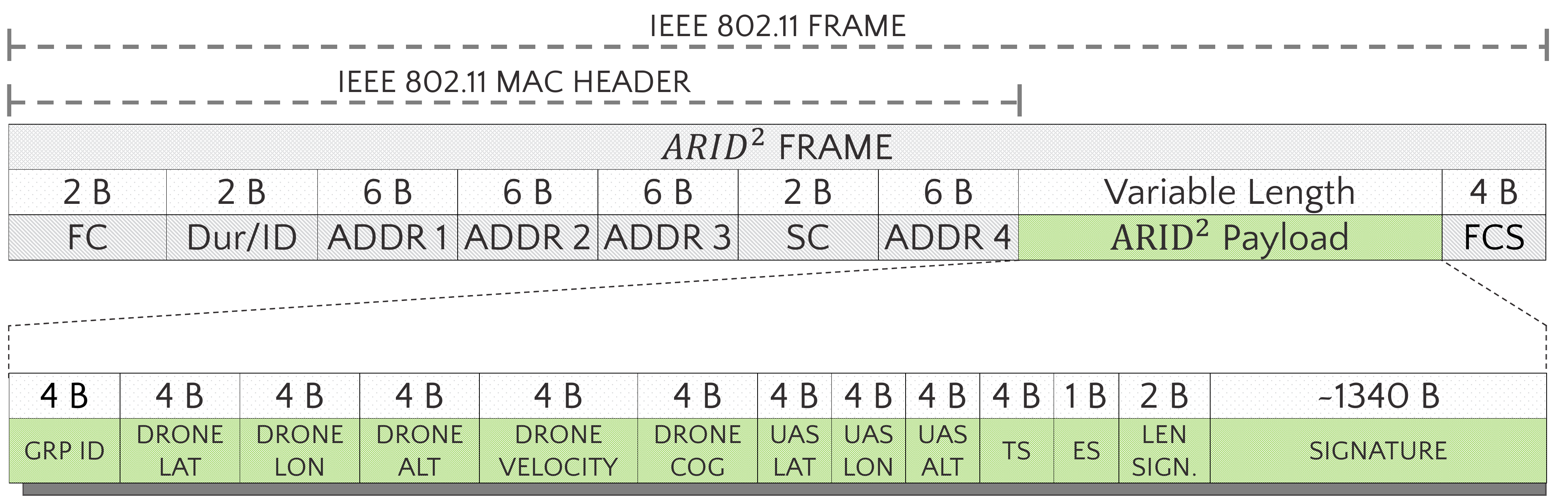}
    \caption{IEEE 802.11b custom Frame and \proto\ Payload Notation.}
    \label{fig:arid2_frame}
\end{figure}

We report the structure of the IEEE 802.11b MAC-layer raw frame (customized for \proto) in Fig.~\ref{fig:arid2_frame}, while Tab.~\ref{tab:messages} provides additional configuration details. Note that, we customized the \ac{PDU} with dedicated a IEEE 802.11 Message ID (\texttt{0xa21d}). \textcolor{black}{Fields such as the Group ID, dependent on the deployment of the scheme, can be enlarged or reduced as necessary.}

\begin{table*}[htbp!]
\caption{IEEE 802.11b MAC-layer Custom Frame and \proto\ Payload Notation.}
\centering
    \begin{tabular}{|P{2.2cm}|P{2.9cm}|P{9.0cm}|}
    \hline
        \textbf{Name Field} & \textbf{Content/Size} & \textbf{Description} \\ \hline
        FC & \texttt{2 B} & Frame Control. \\ \hline
        Duration/ID & \texttt{2 B} & Payload Length. \\ \hline
        Address 1 & \texttt{FF:FF:FF:FF:FF:FF} & Receiver MAC Address. \\ \hline
        Address 2 & \texttt{00:00:00:00:00:00} & Sender MAC Address. \\ \hline
        Address 3 & \texttt{6 B} & N/A. \\ \hline
        Sequence Control & \texttt{2 B} & Sequence Control Field. \\ \hline
        Address 4 & \texttt{6 B} & N/A. \\ \hline
        Payload & \texttt{1386-1390 B} & \proto\ message. \\ \hline
        FCS & \texttt{2 B} & Frame Check Sequence.\\ \hline
        \multicolumn{3}{|c|}{\textbf{\proto\ Payload}}\\\hline
        $GRP_{ID}$ & \texttt{4 B} & Drone Group ID. \\ \hline
        $D_{LAT}$ & \texttt{4 B} & Drone Latitude. \\ \hline
        $D_{LON}$ & \texttt{4 B} & Drone Longitude. \\ \hline
        $D_{ALT}$ & \texttt{4 B} & Drone Altitude. \\ \hline
        $D_{VEL}$ & \texttt{4 B} & Drone Speed. \\ \hline
        $D_{COG}$ & \texttt{4 B} & Drone Course Over Ground. \\ \hline
        $U_{LAT}$ & \texttt{4 B} & Ground Station Latitude. \\ \hline
        $U_{LON}$ & \texttt{4 B} & Ground Station Longitude. \\ \hline
        $U_{ALT}$ & \texttt{4 B} & Ground Station Altitude. \\ \hline
        TS & \texttt{4 B} & Message Timestamp.\\ \hline
        ES & \texttt{1 B} & Emergency Code.\\ \hline
        $L_{SIG}$ & \texttt{2 B} & Signature Length.\\ \hline
        SIG & \texttt{1343-1347 B} & \proto\ Signature.\\ \hline
    \end{tabular}
    \label{tab:messages}
\end{table*}

For the pairing operations, we used the Pairing-Based Cryptography (PBC) Library v0.5.14~\cite{lynn2013pbc}. We adopted the symmetric pairings over Type-A (supersingular) curves defined in the PBC library with the default parameters, which offers the highest efficiency. In our implementation, $p$ is a $160$-bit Solinas prime, which offers $1,024$ bits of discrete-logarithm security. With this Type-A curves setting in PBC, elements of groups $\mathbb{G}$ and $\mathbb{G}_T$ are represented through $512$ and $1,024$ bits, respectively. We selected the cited curve because it provides a security level of $80$~bits, avoiding message fragmentation (not desirable). In particular, starting from a signature of $27,392$~B, we encoded such a string in \texttt{base58}, reducing it to $1,343\sim1,347$~B, i.e. $95.08\%$ less.\\
As for the supporting cryptographic operations, we used the \pcalgostyle{SHA-1} as the hashing function, and a sound cryptographic \ac{PRNG} (\pcalgostyle{/dev/urandom}) as a source of random bits. Finally, we implemented the generic observer and the Authority as separated processes on a Raspberry Pi.\\
We also released at~\cite{arid2_code} a custom Wireshark dissector plugin for the proposed protocol, allowing observers to identify the customized IEEE 802.11 raw frames of $CS-A^2RID$ immediately.\\
Our signature implementation on the Holybro X500 requires $149.025$~kB of Flash Memory and $3.88$~kB of RAM. We also released the source code of \proto\ as open-source, to allow interested researchers and readers to verify our claims and possibly extend \proto\ with additional features~\cite{arid2_code}. We remark that our implementation leverages popular open-source tools, such as the Ubuntu OS, libtins, MAVSDK, and PBC Cryptography Library, supported by a large variety of commercial \acp{UA}.



{\bf DS Scheme.} We also implemented both the $DS-CPA-A^{2}RID$ and the $DS-CCA2-A^{2}RID$ protocols on a reference constrained platform, i.e., the ESPcopter~\cite{ESPcopter}. The ESPcopter is a wirelessly networkable small-size programmable mini-drone, powered by the microcontroller ESP8266 produced by EspressIf~\cite{ESPcopter}. As such, it features an L106 32-bit RISC microprocessor core running at up to $160$~MHz, a WiFi 802.11 b/g/n module for wireless communications, $16$~MB of Flash, and $160$~kB of RAM, being one of the most constrained commercial platforms where to implement, run, and test our protocols. Starting from the ESPcopter code base available at~\cite{ESPcopter_code}, we imported the MIRACL~\cite{miracl} library for cryptography operations on pairing-friendly curves. We remark that, although pairing operations are not executed on the \ac{UA} during the online phase, the protocols described in Sec.~\ref{sec:ds-scheme} require the usage of pairing-friendly curves, whose implementation cannot be found in other cryptographic libraries for constrained devices. Finally, note that for the implementation of the protocols, we used the curve \emph{BN254}, providing a state-of-the-art security level of 128 bits.\\
As for the $CS-A^2RID$ scheme, also the source code of the implementation of the $DS-CPA-A^{2}RID$ and the $DS-CCA2-A^{2}RID$ protocols on the ESPcopter has been released as open-source at~\cite{arid2_code_Eva}.

\subsection{Experimental Results}
\label{sec:results}
We report here the performance assessment of the $CS-A^2RID$ protocol on the Holybro X500 and the $DS-A^2RID$ protocols on the ESPcopter. \textcolor{black}{All the time measurements have been obtained through the function \texttt{std::chrono::high\_resolution\_clock::now()}, with an accuracy of $1$~ns. Moreover, for all the results, we report the $95$\% confidence interval as obtained through the related Matlab routine.}

\begin{figure}[htbp]
    \centering
    \includegraphics[width=\columnwidth]{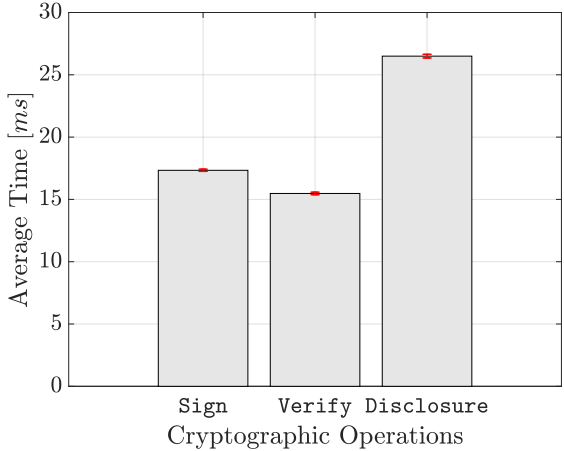}
    \caption{Time required to execute the $CS-A^2RID$ protocol of the \proto\ protocol suite on the HolyBro X500 with mission computer UP Xtreme i7 8665UE.
    }
    \label{fig:crypto}
\end{figure}

\begin{figure}[htbp]
    \centering
    \includegraphics[width=\columnwidth]{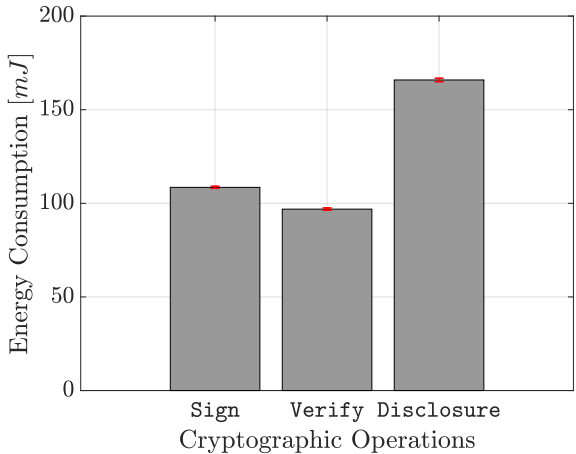}
    \caption{Energy required to execute the $CS-A^2RID$ protocol of the \proto\ protocol suite on the HolyBro X500 with mission computing unit UP Xtreme i7 8665UE.
    }
    \label{fig:energy_operations}
\end{figure}

\begin{figure}[htbp]
    \centering
    \includegraphics[width=\columnwidth]{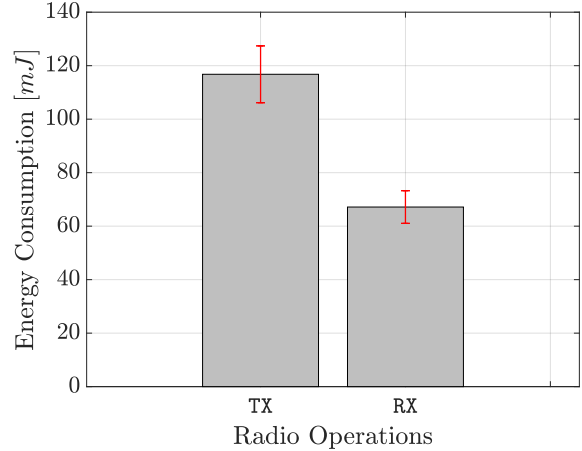}
    \caption{Radio Energy required to transmit and receive \proto\ custom raw IEEE 802.11 frame.}
    \label{fig:energy_radio_operations}
\end{figure}

{$\bm{CS-A^2RID}$.} We report in Fig.~\ref{fig:crypto}, Fig.~\ref{fig:energy_operations}, and Fig.~\ref{fig:energy_radio_operations} the average time and the average energy (due to processing and radio operations) required to execute the cryptographic operations of the $CS-A^2RID$ protocol of the \proto\ protocol suite over $1,000$ tests (with $95$\% confidence intervals).
We notice that the most time-consuming operation is on the \textit{Disclosure Phase}, requiring an average time of $26.498$~ms. The \textit{Signature Generation} procedure in the Online Phase requires $17.335$~ms on average, while the \textit{Signature Verification} requires $15.477$~ms, on average.

As for the energy consumption, to take the measurements, we used a Keysight E36232A DC power supply set, with a voltage of $14.8$~V. Specifically, we measured the difference in the electrical current drained by the drone between two different states: (i) at rest; and, (ii) during the execution of \proto. We computed an average difference of $\approx423$~mA in the electric current drained by the drone over $1,000$~runs, and from such a measure, we extracted energy consumption values.
As reported in Fig.~\ref{fig:energy_operations}, the most energy-consuming operation is always the \textit{Disclosure} procedure, requiring an average energy of $165.888$ mJ. The \textit{Signature Generation} procedure requires $108.525$ mJ on average, while the \textit{Signature Verification} procedure requires $96.895$ mJ.
To estimate the energy consumption of the radio operations on the Holybro X500, we used an Alfa AWUS036ACH card, capable of running in monitor mode and featuring packet injection capabilities. Such a device works with an input voltage of $5$~V, consuming $\approx600$~mA in TX mode and $\approx345$~mA in RX mode, with the IEEE 802.11b protocol~\cite{Keranidis2014} at $2.4$~GHz. We also assumed that a packet is modulated through the standard Direct Sequence Spread Spectrum (DSSS) modulation using Differential Binary Phase-Shift Keying (DBPSK), a Transmission Rate of $1.0$~Mbps on the $22$~MHz channel bandwidth, and a Short Guard Interval of $800$~ns. In this configuration, the $CS-A^2RID$ protocol in \proto\ consumes only $\approx116.7$~mJ per instance in TX mode (i.e., delivered \proto\ frame), and only $\approx67.7$~mJ in RX mode, considering that a frame is transmitted and received in $\approx38.9$~ms. 
Considering that the overall capacity of the battery powering the Holybro X500 drone is $266,400$~J ($5,000$~mAh at $14.8$~V), the $CS-A^2RID$ protocol consumes on average only $\approx 4.073761 \cdot 10^{-5}$\% of the battery of the drone for each instance. Overall, such results prove that the energy consumption of the $CS-A^2RID$ protocol in \proto\ is reasonable, and that anonymous \ac{RID} operations can be executed reliably and in a lightweight fashion, while fully adhering to the \ac{RID} specification.

\begin{figure}[htbp]
    \centering
    \includegraphics[width=\columnwidth]{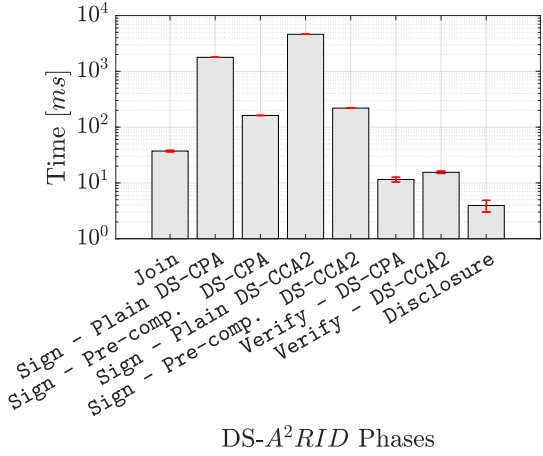}
    \caption{Time required to execute all the phases of the $DS-A^2RID$ protocols of the \proto\ protocol suite, with and without pre-computations, on the ESPcopter and the HP ZBook Studio G5 laptop.
    }
    \label{fig:ds-time}
\end{figure}

{$\bm{DS-A^2RID}$.} We also evaluated the time and energy cost of the two variants of $DS-A^2RID$, namely, $DS-CPA-A^2RID$ and $DS-CCA2-A^2RID$, on the ESPcopter. For the online phase of such protocols, we investigated the performance of two implementation strategies, i.e., without and with pre-computations. The implementation strategy without pre-computations, namely, \emph{plain}, requires the run-time execution of all the modular multiplications and exponentiations detailed in the description of the \emph{Online Phase} in Sec.~\ref{sec:ds-scheme}. Conversely, the implementation strategy with pre-computations requires an offline pre-computation and storage of a number of elliptic curve points correspondent to the ones necessary to run the \emph{Online Phase} of the protocols for a given time. Here, the intuition is that it is possible to trade-off processing time and energy with storage, accelerating the execution of the protocol. We report in Fig.~\ref{fig:ds-time} the execution times of all the phases of the $DS-A^2RID$ protocols, including the two implementation strategies mentioned above. While the online phase has been executed on the ESPcopter, the join, verify, and open phases have been executed on a HP ZBook Studio G5 laptop equipped with two Intel(R) Core(TM) i7-9750H CPUs running at $2.60$~GHz. 

Overall, we notice that the execution of the \emph{plain} versions of the $DS-A^2RID$ protocols requires more than $1$~s (specifically, $\approx4.616$~seconds for the $DS-CCA2-A^2RID$ and $\approx1.784$~seconds for the $DS-CPA-A^2RID$), not allowing to fulfill the time requirement set in the \ac{RID} specification. Conversely, as modular multiplication and exponentiation operations are the most time-consuming ones, through pre-computations, the ESPcopter can complete the $DS-CCA2-A^2RID$ in $0.22$~seconds and the $DS-CPA-A^2RID$ in $0.16$~seconds, definitely fulfilling the above-mentioned requirement.

\begin{figure}[htbp]
    \centering
    \includegraphics[width=.8\columnwidth]{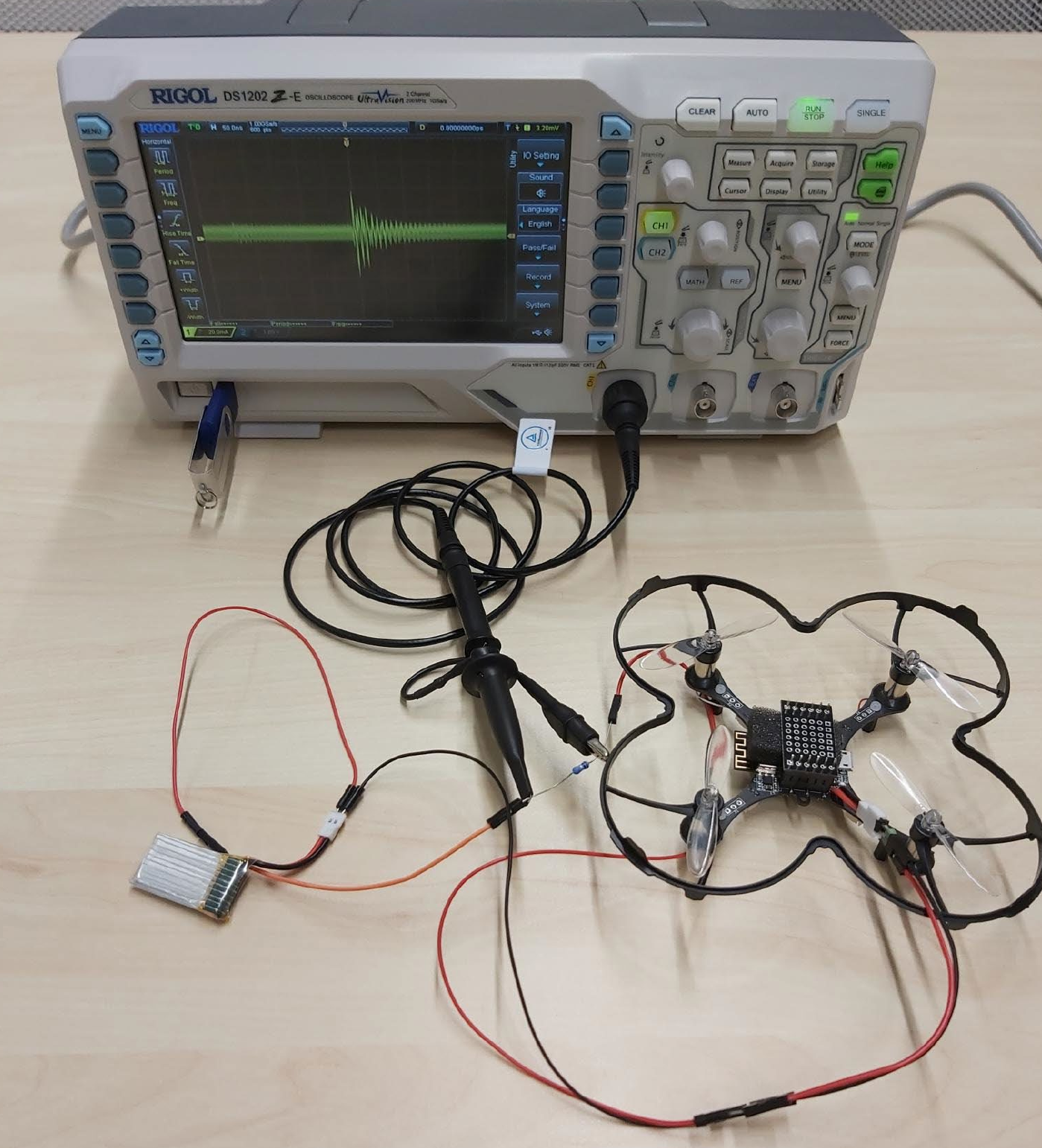}
    \caption{Experimental setup used for acquisition of energy consumption measurements.}
    \label{fig:energy-setup}
\end{figure}

We also measured the energy consumption required by the signing operations on the ESPcopter, both in the \emph{plain} mode and with pre-computations. Fig.~\ref{fig:energy-setup} shows the testbed used for energy consumption measurements.

\begin{figure}[htbp]
    \centering
    \includegraphics[width=\columnwidth]{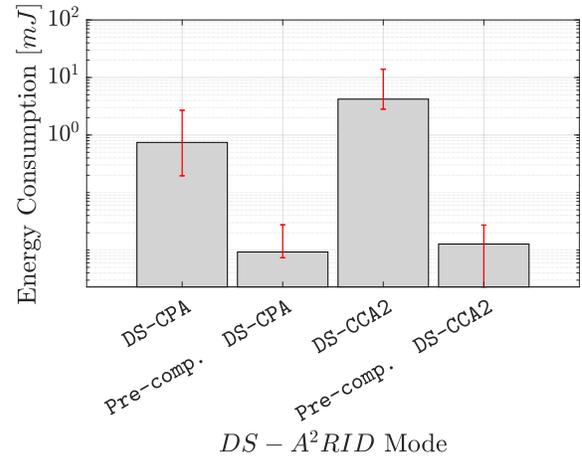}
    \caption{Energy required to generate anonymous \ac{RID} messages through $DS-A^2RID$ protocols on the ESPcopter, with and without pre-computations.}
    \label{fig:ESPcopter_energy}
\end{figure}

\textcolor{black}{We used the oscilloscope RIGOL DS1202Z-E, acquiring samples with a real-time sample rate of up to $1$~Gsa/s on two parallel channels~\cite{rigolshop}}. As the oscilloscope measures voltage only, in line with the methodology used in~\cite{sciancalepore2018_sensors}, we sampled the voltage drop to the terminals of a shunting resistor of $0.1~\Omega$, bridging the pins in series with the chipset of the ESP8266 and the battery powering the ESPcopter. We acquired several runs of each protocol, and we post-processed the acquired samples through Matlab. Fig.~\ref{fig:ESPcopter_energy} shows the results of our analysis.

In line with the time measurements, using pre-computations reduces the energy consumption significantly. Looking at the $DS-CPA-A^2RID$ solution, using pre-computed values requires only $9.25$~$\mu$J per message signing, on average, down $98.75$~\% from the $0.74$~mJ required for the same scheme, without pre-computations. At the same time, using $DS-CCA2-A^2RID$ with pre-computations requires $12.74$~$\mu$J, $37.73$~\% more than $DS-CPA-A^2RID$ with pre-computations. Considering that the ESPcopter is powered by a tiny Li-Po battery with a capacity of $260$~mAh, signing a message with $DS-CPA-A^2RID$ requires only $9.63 \cdot 10^{-6}$~\% of the overall available energy, introducing very limited energy overhead.

\begin{figure}[htbp]
    \centering
    \includegraphics[width=\columnwidth]{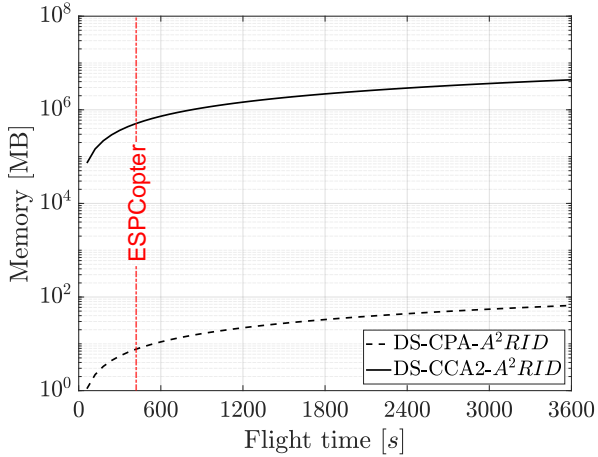}
    \caption{Memory space required to run the $DS-CCA2-A^2RID$ and the $DS-CPA-A^2RID$ protocols, with different flight times, assuming to deliver exactly one \ac{RID} message per second.}
    \label{fig:ds-memory}
\end{figure}

At the same time, we highlight that the price to pay for the acceleration of signature operations is on the memory requirements of the protocol, which increase with the amount of information to be stored on the \ac{UA} and thus, on the duration of the flight. Figure~\ref{fig:ds-memory} shows the memory space required to pre-compute the values necessary to run both the $DS-CCA2-A^2RID$ and the $DS-CPA-A^2RID$ protocols, with different flight times, assuming to deliver exactly one \ac{RID} message per second, i.e., the minimum requirement to adhere to the \ac{RID} rule. 

As the $DS-CCA2-A^2RID$ protocol requires more modular multiplications and exponentiation operations at run-time, it requires more memory than the $DS-CPA-A^2RID$. For a flight time of $420$~seconds, $DS-CCA2-A^2RID$ requires $510,049$~MB, while $DS-CPA-A^2RID$ requires $7.70$~MB only, achieving a consistent reduction.

Overall, these results demonstrate that, using pre-computations, it is possible to achieve anonymous direct authentication and remote identification of commercial \acp{UA}, while fulfilling all the requirements imposed by the \ac{RID} regulation, at the cost of a variable memory increase. In this context, we highlight that this is a reasonable trade-off for commercial drones. Indeed, while processing capabilities require expensive and possibly bulky hardware, with more significant energy consumption and bulkier batteries, increasing memory size typically requires only attaching a bigger SD card, which is less energy-demanding and lightweight to carry around.

\subsection{Comparison}
\label{sec:comparison}

In this section, we compare the performance of the protocols in the \proto\ protocol suite against a benchmarking solution, i.e., the legacy version of the ARID protocol in~\cite{tedeschi2021_acsac}. Note that, to the best of our knowledge, as summarized in Sec.~\ref{sec:related}, the protocol in~\cite{tedeschi2021_acsac} is the only current solution providing anonymity for RemoteID-compliant drones, being the straightforward choice for our comparison. 

For the comparison, we implemented and ran the \emph{Online Phase} of all the protocols on a common platform, i.e., a Dell XPS 9570 laptop, integrating an Intel(R) Core(TM) i7-8750H CPU running at $2.20$~GHz, and equipped with the Ubuntu 22.04 LTS operating system and $32$~GB of RAM. We highlight that the hardware selected for comparison is very similar to the ones available on the Holybro X500 drone. At the same time, running all the protocols on a constrained platform such as the ESPcopter is not possible, as such a platform (and similar ones) cannot run processing-intensive approaches. 

\begin{figure}[htbp]
    \centering
    \includegraphics[width=\columnwidth]{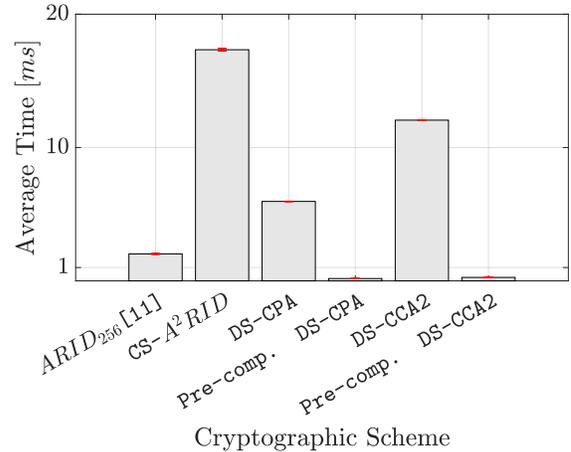}
    \caption{Time required for anonymous signature generation in the protocols of the $A^2RID$ suite and the approach in~\cite{tedeschi2021_acsac}.}
    \label{fig:comparison}
\end{figure}

Figure~\ref{fig:comparison} reports the average time (and related $95$\% confidence intervals) required to generate anonymous RID messages on the mentioned hardware platform, using the approach in~\cite{tedeschi2021_acsac}, the $CS-A^2RID$ protocol (Sec.~\ref{sec:cs-scheme}), and the $DS-A^2RID$ protocols (Sec.~\ref{sec:ds-scheme}), without and with pre-computations. Note that, for all the protocols excluding $CS-A^2RID$, we considered the configurations providing the same security level, i.e., $256$~bits. For $CS-A^2RID$, instead, we considered a security level of $80$~bits.

We notice that, although being less secure ($80$ bits instead of $256$), the $CS-A^2RID$ protocol is the most time-consuming solution, taking on average $17.34$~msec to generate anonymous RemoteID messages. This result is consistent with the cryptography operations required by such a protocol, involving bilinear pairing operations during anonymous signature generation. At the same time, when considering the respective memory-friendly versions, although not relying on bilinear pairing operations at signature generation time, the two $DS-A^2RID$ protocols are more time-consuming than the approach in~\cite{tedeschi2021_acsac}, requiring $12.05$~msec ($DS-CCA2-A^2RID$) and $5.95$~msec ($DS-CPA-A^2RID$), respectively, against the $2.02$~msec of the approach in~\cite{tedeschi2021_acsac}. This additional time (and energy) overhead is the price to pay to provide direct authentication, instead of the brokered authentication provided by~\cite{tedeschi2021_acsac}. However, by adopting pre-computations, the $DS-A^2RID$ protocols trade off computation time with memory storage, allowing to significantly reduce the time to generate anonymous RemoteID-compliant messages at the cost of increased memory footprint. In particular, by adopting pre-computations, the $DS-CCA2-A^2RID$ and $DS-CPA-A^2RID$ protocols take on average only $0.26$~msec and $0.17$~msec. We finally highlight that, compared to the solution in~\cite{tedeschi2021_acsac}, the protocols in the $A^2RID$ suite also allow direct authentication of the messages, with significant advantages in terms of ease of deployability and maintenance.

\section{Conclusion}
\label{sec:conclusion}
In this paper, we proposed \proto, a protocol suite for anonymous direct authentication and remote identification of \aclp{UA}. By adopting the protocols in \proto, \acp{UA} can maintain anonymity while broadcasting standard-compliant \acl{RID} messages. At the same time, on reporting by observers, the \acl{USS} can recover the identity of misbehaving \acp{UA}, for follow-up actions. 
The protocols in \proto\ are characterized by different requirements, being able to run even on very constrained \acp{UA}, with minimal processing, storage, and energy load. As a reference, through smart pre-computations, the most lightweight protocol, namely $DS-CPA-A^2RID$, can provide CPA-full anonymity for constrained \acp{UA} while requiring only $0.16$~seconds for signature generation on the ESPcopter, as well as a minimal energy toll. Thanks to the native integration into the standardized \emph{drip} networking architecture, \proto\ enjoys a straightforward adaptation into actual deployments, contributing to enforce drones' anonymity and invasion accountability. Finally, we highlight that we also released the source code of \proto\ as open-source, to allow interested researchers and industry to verify our claims and possibly extend \proto\ with additional features~\cite{arid2_code}~\cite{arid2_code_Eva}, as well as to check the viability of further research directions. \textcolor{black}{Future works include evaluating the feasibility of \proto\ for improving location privacy of \acp{UA}, as well as assessing its feasibility for mitigating \acp{UA} tracking. We also plan to evaluate the feasibility of selective online disclosure of UA identity to specific parties.
}

\section*{Acknowledgements}
The authors would like to thank the anonymous reviewers, that helped improving the quality of the paper. This work has been partially supported by the INTERSECT project, Grant No. NWA.1162.18.301, funded by Netherlands Organisation for Scientific Research (NWO).
This publication was also supported by Technology Innovation Institute - Abu Dhabi, and awards NPRP-S-11-0109-180242 from the QNRF-Qatar National Research Fund, a member of The Qatar Foundation, and NATO Science for Peace and Security Programme - MYP G5828 project ``SeaSec: DronNets for Maritime Border and Port Security''. The findings reported herein are solely responsibility of the authors.

\bibliographystyle{IEEEtran}
\bibliography{references}

\end{document}